%% file: main.tex
\documentclass[preprint,authoryear]{elsarticle}
\usepackage[T1]{fontenc}
\usepackage[utf8]{inputenc}
\usepackage{amsmath, amssymb, amsthm}
\usepackage{mathtools}
\usepackage{bm}
\usepackage{graphicx}
\usepackage{booktabs}
\usepackage{array}
\usepackage{tabularx}
\usepackage{multirow}
\usepackage{subcaption}
\usepackage{hyperref}
\usepackage{url}
\usepackage{xcolor}
\usepackage{enumitem}
\usepackage{dsfont}

\makeatletter
\def\ps@pprintTitle{%
  \let\@oddhead\@empty
  \let\@evenhead\@empty
  \def\@oddfoot{\centerline{\thepage}}%
  \let\@evenfoot\@oddfoot}
\makeatother
\AtBeginDocument{\hypersetup{
    colorlinks=true,
    linkcolor=black,
    citecolor=black,
    urlcolor=blue!50!black,
}}

\newtheorem{proposition}{Proposition}
\newtheorem{corollary}{Corollary}

\theoremstyle{definition}

\theoremstyle{remark}

\newcommand{\RV}{\mathrm{RV}}
\newcommand{\Var}{\mathrm{Var}}

\newcommand{\E}{\mathbb{E}}

\newcommand{\Hhat}{\widehat{H}}
\newcommand{\hHat}{\widehat{H}}

\begin{document}

\begin{frontmatter}

\title{Rough Volatility Across Assets}

\author[unh]{Saad Mouti}
\ead{smouti@newhaven.edu}

\address[unh]{Department of Mathematics and Physics, University of New Haven, West Haven, CT, USA}

\begin{abstract}
\noindent
We measure volatility roughness across asset classes using a common data infrastructure and pipeline. Our data covers 3,926 United States equities, 34 CME futures roots, rates, FX, and commodities, and options on 44 underlyings over 2010-2025. Realized volatility is rough everywhere. The class-median Hurst estimate ranges from $0.05$ (livestock) through $0.07-0.10$ (rates, FX, agriculture, energy, metals) to $0.13$ (single stocks) and $0.20$ (equity indices). The option-implied measure identifies $H$ only where the leverage effect produces a clean skew term structure. For the equity indices, implied estimates of $0.21-0.28$ are just above realized volatility $H$, while for rates and FX the ATM skew regression fails with an R-squared near zero even though realized volatility remains rough. We also show a mean-reversion contamination formula for the second-moment estimator of the roughness for the stationary fractional Ornstein-Uhlenbeck process. The local slope of the increment second moment deviates from $2H$ by $(1-H)\Gamma(2H+1)(\kappa\Delta)^{2-2H}$ for all $H\in(0,1)$. A correction framework for the second moment, when the log realized volatility measure has additive noise, raised the $H$ estimate slightly but nowhere near the Brownian diffusion framework. Finally, a failure taxonomy discusses where rough-volatility methods apply and where they fail, suggesting alternative paths to explore the rough volatility paradigm further.
\end{abstract}


\end{frontmatter}

%% ============================================================
%% Sections
%% ============================================================

\input{sections/01_introduction}

\input{sections/02_literature}
\input{sections/03_data}
\input{sections/04_methodology}
\input{sections/05_simulation}
\input{sections/06_results_equities}
\input{sections/07_results_cross_asset}
\input{sections/08_failure_taxonomy}
\input{sections/09_conclusion}

%% Bibliography
\bibliographystyle{elsarticle-harv}
\bibliography{references}

%% Appendices
\appendix
\input{sections/A1_mrc_proofs}

\input{sections/A4_estimator_formulas}

\end{document}

%% file: sections/01_introduction.tex
\section{Introduction}
\label{sec:introduction}

Since the introduction of the rough volatility model by \citet{gatheral2018volatility}, which documented that equity-index log volatility scales like fractional Brownian motion with Hurst exponent $H$ of order $0.1$, this model has become a standard modeling framework. On the one hand, it explains the short-maturity skew \citep{alos2007short, fukasawa2017short, bayer2016pricing} and admits tractable pricing when coupled with a Heston diffusion \citep{elEuch2019characteristic}. However, its empirical evidence remains concentrated in equity, with the benchmark coming from equity indices. This paper estimates $H$ across asset classes with 3,926 U.S. equities, 34 CME futures roots covering equity indices, rates, FX, energy, metals, agriculture, and livestock, and options on 44 underlying assets.

Realized volatility is rough in every asset class. Class medians range from $0.05$ (livestock) through $0.08$ to $0.10$ (rates, FX, agriculture, energy, metals) to $0.13$ (single stocks) and $0.20$ (equity indices), with regression fits near $R^2=0.98$. The ATM skew identifies $H$ only where the skew term structure is clean. For equity indices, the implied volatility-based $H$ is $0.21$ to $0.28$, slightly above the realized volatility-based $H$. For rates and FX the skew regression fails, with an R-squared near $0$, while realized volatility remains rough. For seasonal commodities, the implied estimate is also fragile, partly because of the maturity effect in futures.

Assuming monofractality, our primary estimator is ordinary least squares on the log second moment of log-realized variance increments. We present a proposition on the mean-reversion contamination of fractional Ornstein-Uhlenbeck and deduce that using a lag of up to ten days avoids this effect. We also explore how realized volatility measurement errors affect the estimation of $H$ and address this with seven different estimators, an explicit offset correction, quality checks, and one-second-based estimators for the most liquid assets. Each refinement slightly raises the equity estimates, and the most liquid subset median saturates at $0.15$. A simulation study validates the framework.

The key contributions of this paper are the following: (i) a single pipeline estimate of volatility roughness across the exchange-traded universe; (ii) a mean-reversion contamination formula for the $H$ estimator from realized volatility and a measurement of the bias induced; (iii) a comparison of the implied vs. realized volatility-based estimation across 41 assets; and (iv) a taxonomy of where option-based identification fails (rates and FX, seasonal underlyings). On the equity panel itself, \citet{bennedsen2022decoupling} already estimate roughness for thousands of U.S. stocks; we are not adding breadth of the panel, but instead explore other classes and the two Hurst exponent channels. Section \ref{sec:literature} reviews the literature, Sections \ref{sec:data} and \ref{sec:methodology} present the data and the methods, Section \ref{sec:simulation} the simulation, Sections \ref{sec:results:equities} and \ref{sec:results:cross} the results, and Section \ref{sec:failure} the failure taxonomy. Finally, Section \ref{sec:conclusion} concludes.

%% file: sections/02_literature.tex
\section{Related Literature}\label{sec:literature}

While prior literature on diffusion models for volatility assumed trajectories with regularity close to that of Brownian motion, empirical observations suggest volatility is a long-memory process. As such, various authors proposed models that allow a wider range of regularity than Brownian diffusion. Among the first works in this direction is \citet{comte1998long}, which suggested a mean-reverting process using fractional Brownian motion with Hurst parameter $H>\frac{1}{2}$. This makes the volatility non-Markovian, ensuring long memory with persistence. A large body of literature has since developed around such fractional volatility models, such as \citet{cheridito2003fractional}. However, the modern rough-volatility paradigm began with \citet{gatheral2018volatility}, who showed that the log realized volatility of equity indices scales like fractional Brownian motion with $H$ of order $0.1$ and proposed the rough fractional stochastic volatility, a fOU process with a fractional Brownian motion with small $H$ and slow mean reversion. On the pricing side, \citet{bayer2016pricing} demonstrated that the rough Bergomi model reproduces the SPX skew term structure with a handful of parameters, and \citet{elEuch2019characteristic} derived the characteristic function of the rough Heston model. \citet {elEuch2018microstructural} studied microstructure foundations, roughness, and leverage emerging from order-flow dynamics.

\citet{gatheral2018volatility} used regression of the log absolute moments of log realized variance on log lag to estimate the Hurst exponent. \citet{bolko2023gmm} built a GMM estimator that treats realized variance as a noisy proxy; \citet{wang2023modeling} model log realized variance directly as an fOU process with general $H$ and estimate it based on the ratio of two second-order differences of observations from different frequencies and the method of moments. \citet{garcin2022long} document the dilemma between short and long time scales in fitting fOU models, while \citet{fukasawa2022consistent} construct a noise-robust likelihood estimator and recover roughness under high-frequency asymptotics. \citet{livieri2018rough} estimate $H$ using implied volatility proxies of realized volatility as the option reaches its expiration and find an $H$ of order $0.3$ for S\&P 500. \citet{mouti2023rough} documents rough volatility from range-based volatility estimators on exotic indices and stocks and find that $H$ is close to zero. \citet{zarhali2025volatility} find that single stocks are much rougher than indices. \citet{bennedsen2022decoupling} estimate the roughness for thousands of U.S. equities in a model that decouples short and long-term volatility behavior, and is the closest empirical predecessor of our paper. From the implied volatility side, \cite{alos2007short} and \citet{fukasawa2017short} provides the option-side identification through the short maturity at-the-money skew, $\psi(T) \sim T^{H-1/2}$.

Outside equities, the evidence is thinner and more recent. \citet{alfeus2022forecasting} find long-memory in commodity volatility forecasting, and \cite{alfeus2024implied} calibrate a rough volatility model for oil options to the volatility surface and find that a small Hurst parameter fits better than $H=1/2$. \citet{daluiso2026rough} develop a general rough volatility model for commodities that provides an automatic calibration of the initial term structure of the futures prices and an appropriate treatment of the Samuelson effect, and estimate $H$ between $0.14$ and $0.21$ from the realized volatility of WTI futures.

%% file: sections/03_data.tex
\section{Data}\label{sec:data}
\subsection{Asset universe and data sources}\label{sec:data:universe}
This paper uses data across four asset classes (equities, rates, foreign exchange, and commodities), along with options on a subset of the underlyings, all via Databento. Equity data (including ETFs) come from the Nasdaq TotalView-ITCH feed. Futures on Treasury rates, foreign exchange, equity indices, and commodities are extracted from CME Globex MDP 3.0. Option data come from two sources: CME MDP 3.0 for options on futures and OPRA Pillar for options on common stocks, ETFs. and indices. The implied volatility covers 44 underlyings, 33 CME option roots, and 11 from OPRA. Table \ref{tab:data} summarizes the sources.

Equity data consist of 1-minute OHLC prices of NASDAQ-listed stocks and span May 2018 to December 2025. We exclude stocks with fewer than 500 consecutive trading days and with zero volume. Futures data come from the CME Globex MDP 3.0 feed from June 2010. Additionally, we extract one-second frequency of the forty most liquid equity stocks, drawn from the same Nasdaq feed. Option settlements and closes come from the CME statistics schema (official cleared settlement marks, available for every listed strike) and from OPRA.

\begin{table}[t]
\centering
\caption{Data summary by asset class.}
\label{tab:data}
\footnotesize
\setlength{\tabcolsep}{4pt}
\resizebox{\textwidth}{!}{%
\begin{tabular}{lllll}
\toprule
Asset class & Instruments & Source & Sample period & Type \\
\midrule
Equities        & 3,926 Nasdaq stocks         & XNAS.ITCH + BASIC & 2018-05--2025-12 & 1 min \\
Liquid subset   & 40 stocks and ETFs            & XNAS.ITCH         & 2018-05--2025-12 & 1 s \\
Equity indices  & ES, NQ, YM                    & CME GLBX.MDP3     & 2010-06--2025-12 & 1 min \\
Rates           & ZT, ZF, ZN, ZB, GE            & CME GLBX.MDP3     & 2010-06--2025-12 & 1 min \\
FX              & 6A--6S (8 roots)              & CME GLBX.MDP3     & 2010-06--2025-12 & 1 min \\
Commodities     & 18 roots                      & CME GLBX.MDP3     & 2010-06--2025-12 & 1 min \\
Options (CME)   & 33 futures option roots       & CME statistics    & 2010-06--2025-12 & settle \\
Options (OPRA)  & SPX, XSP, NDXP, SPY,+ 7 single stocks  & OPRA              & 2013-04--2025-12 & close \\
\bottomrule
\end{tabular}}
\end{table}

\subsection{Data construction}\label{sec:data:construction}

The equity universe consists of 3,926 Nasdaq-common stocks and ETFs which have at least 120 usable one-minute data points from 2018 to 2025; each trading day is assigned to the regular-session grid of 390 one-minute OHLC prices, missing prices are forward-filled and are noted as such; each symbol/day has a fill fraction in order to ensure that the results reflect the conditional data quality; in order to be included in the quality subset, an asset must have at least 500 estimation days and for the median day at least 80\% of the fill fraction; this results in 1,409 assets; the longest equity history available is 1,929 trading days; realised variance for equities is calculated using intraday prices and does not take overnight returns into account; the same rule applies to futures, even though they have a shorter overnight trading halt; the universe also includes companies that delist or convert during the sample period; this involves 428 assets that exit before the end of the sample.

Futures are continuous front-month series built with the volume-based roll rule. The series switches to the next contract when its traded volume over-takes the expiring one, so it follows the most liquid contract. %This is a download feature that Databento allows. 
Estimation is performed on the front-month series, where liquidity is concentrated. Roll dates contribute one increment out of roughly 4,100 and have no measurable effect on the second moment. 

Option chains are created using the daily definitions and either the settlement or closing prices. About 190,000 asset-days are converted into implied volatilities. The discount rate and forwards are derived from put-call parity. In the case of CME options, the futures settlement serves as the forward, and for each (date, expiry) pair, a regression of $(C - P)$ on $(F-K)$ yields the discount factor. For equity and index options, both the discount factor and the forward are unknown. A regression of $(C - P)$ on $K$ gives the discount factor from the slope and the forward from the intercept. Implied volatilities are calculated using the Black-76 formula for futures options and for European index options, and the Barone-Adesi-Whaley (BAW) method for American single-stock options. Although CME futures options are of the American style, the early-exercise premium is very small near the money (for example, in the case of American-style ES options, using BAW causes the median implied volatility to change by 0.02 volatility points), so for all futures options near at-the-money (and since our analysis is focused on this area) the Black-76 formula is applied. For each (date, expiry) combination, the at-the-money skew is the slope of an ordinary least-squares regression of implied volatility on log-moneyness k within the range $|k| \leq 0.08$, with at least 10 strikes required (for CME) or 20 strikes with positive volume (for OPRA). The monthly and weekly SPX roots are combined, retaining the one with higher volume for the same expiries. The time window used for the skew regressions extends from 14 days to 1 year; below two weeks, the skew is dominated by jumps and the data from expiry weeks are not reliable. 

%% file: sections/04_methodology.tex
\section{Methodology}\label{sec:methodology}

We assume that the asset log-price $S_t$ follows a continuous semimartingale
\begin{align*}
dS_t = \mu_tdt + \sigma_t dW_t
\end{align*}
where $\mu_t$ is the drift, $W_t$ a one-dimensional Brownian motion, and $\sigma_t$ the volatility process. The rough volatility framework rests on two stylized facts: (a) the shape of the at-the-money implied volatility skew, and (b) the roughness of the realized volatility time series. We adopt this framework and develop estimators for $H$ from both channels.

\citet{gatheral2018volatility} explored the q-th absolute moments of the increments of the log-volatility, i.e., $M_q(\Delta):= \mathbb{E}[|\log(\sigma_{\Delta}) - \log(\sigma_0)|^q]$, estimated by their empirical counterpart
% Assuming the increments of the volatility process are stationary and that the law of large numbers can be applied, those q-th absolute moments can be estimated by their empirical counterpart:
\begin{align}\label{eq:m(q,delta)}
m(q, \Delta) = \frac{1}{N}\sum_{k=0}^{N-1}|\log(\sigma_{(k+1)\Delta})-\log(\sigma_{k\Delta})|^q.
\end{align}
under the assumption that $m(q,\Delta)\approx K_q\Delta^{qs_q}$ as $\Delta \rightarrow 0$ for some $s_q>0$ and $K_q>0$ with $s_q$ read as the regularity of the volatility in $\ell_q$ norm. For fractional Brownian motion with Hurst parameter $H$, $s_q=H$ for every $q$, the defining property of a monofractal process. Using Oxford-Man realized library for the S\&P 500 and NASDAQ indices, they find $s_q$ independent of $q$ and between 0.08 and 0.15, consistent with fractional Brownian scaling with $H < 1/2$. On the implied volatility side, \citet{alos2007short} and \citet{fukasawa2017short} show that the at-the-money skew behaves as the power law $\psi(T) \sim T^{H-1/2}$ at short maturities, so $H$ can be estimated from option prices independently of any realized-volatility measurement. Sections \ref{sec:method:realized} and \ref{sec:method:implied} develop the two estimators. Sections \ref{sec:method:correction} and \ref{sec:method:mrc} discuss shortcomings of second-moment estimation and explore some solutions, while Section \ref{sec:method:rv} introduces the realized volatility measures used in the paper.

\subsection{Hurst parameter from realized volatility}\label{sec:method:realized}
%\subsubsection{Second moment}\label{sec:method:m2}

Starting from the monofractal evidence, we focus on the second absolute moment, setting $q=2$ in \ref{eq:m(q,delta)}. Let $\text{RV}t$ denote a daily realized-variance estimate and $X_t=\log(\text{RV}t)$ its logarithm. For fractional Brownian motion with Hurst parameter $H$, the second moment of the increments obeys

\begin{align}\label{eq:m2_scaling}
M_2(\Delta) = E[|\log(\sigma_{t+\Delta}) - \log(\sigma_{t})|^2] = \nu^2 \Delta^{2H},
\end{align}
where $\nu$ is the vol-of-vol, so that the $\log(M_2(\Delta))$ is affine in $\log(\Delta)$ with slope $2H$. The baseline estimator replaces $M_2$ with its sample counterpart $m(2,\Delta)$ of \eqref{eq:m(q,delta)} at lags $\Delta = 1, \ldots, \Delta_{\max}$ and $\hat{H}$ is half the ordinary least-squares slope of $\log(m(2,\Delta))$ on $\log(\Delta)$. We report the regression $R^2$ with every estimate. Empirical fits are typically excellent ($R^2\approx 0.98$). The regression errors are correlated across lags, so the $R^2$ measures the quality of the power-law shape, but doesn’t report estimation precision. Precision is quantified by the Monte Carlo dispersion of the estimator at rough $H$.

Figure \ref{fig:m(2,delta):1} shows the object being fitted, for SPY and the ES front-month series (TSRV). The log-log plot is linear on the short-lag window, with lag-10 estimates of $0.186$ (SPY) and $0.201$ (ES). Extending the window to 40 lags lowers the fit to $0.161$ and $0.162$, and on lags 40 to 250 the curve flattens toward slopes of $0.076$ and $0.092$. Section \ref{sec:method:mrc} quantifies this behavior, which is due to mean reversion, and a lag-restricted window is designed to avoid its effect. So our primary specification uses $\Delta_{\max}=10$ trading days, for which the mean-reversion bias is small (see Section \ref{sec:simulation}).

\begin{figure}[ht]
\centering
\includegraphics[scale=0.6]{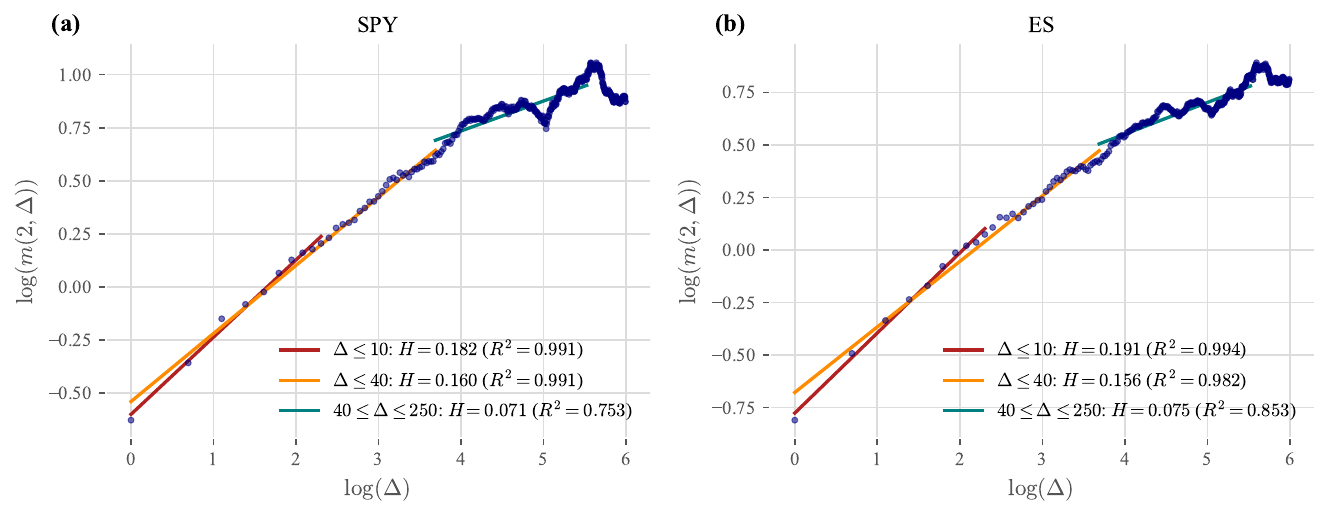}
\caption{$\log(m(2,\Delta))$ as a function of $\log(\Delta)$ with OLS fits on different window lags $\Delta \leq 10$, $\Delta \leq 40$ and $40 \leq \Delta \leq 250$; (a) SPY, May 2018 to December 2025, (b) ES front-month, June 2010 to December 2025}
\label{fig:m(2,delta):1}
\end{figure}

\subsection{Realized variance estimators}\label{sec:method:rv}
The volatility process is unobserved, so we estimate the Hurst parameter from realized measures of volatility computed from intraday data. We compute daily realized variance from one-minute log returns with seven estimators, tuned throughout to their literature specifications, in the following order (more detailed implementation formulas in \ref{app:realized}):
\begin{enumerate}
\item[(i)] $\text{RV}^{5m}$: realized variance from five-calendar-minute subsampled returns \citep{liu2015does};
\item[(ii)] RK: Realized kernel with Parzen weights and the data driver bandwidth of \citep{barndorff2009realized};
\item[(iii)] TSRV: the two-scale estimator of \citep{zhang2005tale}, with the slow scale fixed at five minutes;
\item[(iv)] PAV: pre-averaging (\citep{jacod2009microstructure}) with window $k_n = \lceil \theta\sqrt{n}\rceil$, $\theta=1$;
\item[(v)] PABPV: pre-averaged bipower variation, robust to both noise and jumps;
\item[(vi)] BPV: bipower variation \citep{barndorff2004power}, robust to jumps;
\item[(vii)] C-TRV: the threshold estimator of \citet{mancini2009non} in the corrected form of \citet{corsi2010threshold}, with threshold $\vartheta = 9\hat\sigma^2$, $\hat{\sigma}^2 = \mathrm{BPV}/n$, and exceedances replaced by their conditional expected value.
\end{enumerate}
TSRV is the primary estimator because it is noise-robust, requires no bandwidth tuning beyond the five-minute slow scale, and behaves well in simulation. RK and PAV serve as noise-robust cross-checks; BPV and C-TRV isolate the contribution of price jumps.

\subsection{Correction to the realized estimator}\label{sec:method:correction}
Two opposing biases separate $\Hhat$ from the roughness of the spot path. A realized measure estimates the integrated variance $\int_t^{t+1}\sigma_u^2du$, so the regression runs on a daily average of the variance path rather than on the spot variance, and averaging biases the slope upward. For a spot log-volatility that is exactly fBm, the second moment of its $\delta$-averaged path is $\nu^2\Delta^{2H}f(\delta/\Delta)$ with
\begin{align*}
f(\theta) = \frac{(1+\theta)^{2H+2} - 2 - 2\theta^{2H+2} + (1-\theta)^{2H+2}}{\theta^{2}(2H+1)(2H+2)},
\end{align*}
the smoothing function of \citet[Appendix C]{gatheral2018volatility}. Measurement error acts in the opposite direction, through a constant offset discussed below. The estimand of this paper is therefore the roughness of the integrated volatility, the object reported throughout the empirical literature. Since averaging can only smooth, the spot path is at least as rough as the estimates indicate.

In order to make the offset explicit, we treat the error as additive on the log scale, $\log(\hat\sigma_t) = \log(\sigma_t) + \epsilon_t$, where $\epsilon_t$ are serially uncorrelated, independent of the volatility path, and of variance $\omega^2$. By writing $M_2(\Delta)$ for the second moment of the true log-volatility, the measured series has second moment $M_2(\Delta) + 2\omega^2$ at every lag, with local slope
\begin{align*}
\frac{\partial \log(M_2(\Delta) + 2\omega^2)}{\partial \log(\Delta)} = 2H \frac{M_2(\Delta)}{M_2(\Delta) + 2\omega^2} < 2H.
\end{align*}
The slope is attenuated at every lag and mostly at short lags and small $H$. A flatter power law is harder to distinguish from a constant, so the correction matters most in the rough regime.

The offset is removed with a two-estimator correction, which pairs a second realized measure $b$ with $a = \RV^{\mathrm{5m}}$. The two measured estimates share the true $M_2(\Delta)$, so their second moments differ by the constant $2(\omega_a^2-\omega_b^2)$ independent of the lag, and the mean across lags of the empirical difference $m^{(a)}(2,\Delta)-m^{(b)}(2,\Delta)$, denoted $\delta_m$, estimates the noise-variances difference. The offset is found by a joint fit of the first increment autocovariance $\gamma(k)$ with $\gamma(k) = \frac{\nu^2}{2}(|k+1|^{2H} - 2|k|^{2H}+|k-1|^{2H})$ of the fractional Gaussian noise of the one-day increments of the true log-volatility, additive noise shifts $\gamma(0)$ by $2\omega^2$ and $\gamma(1)$ by $-\omega^2$ and leaves $k\geq 2$ unchanged, so we minimize over $(H, \nu, \omega_a)$
\begin{align}\label{eq:offset-fit}
\sum_{k=0}^{2}\left[\hat\gamma_a(k)-\gamma(k;H,\nu,\omega_a)\right]^2
+\left[\hat\gamma_b(k)-\gamma(k;H,\nu,\omega_b)\right]^2,
\end{align}
subject to $\omega_a^2 = \omega_b^2 + \delta_m/2$. Lags beyond $k=2$ add no information because the noise enters only the first two autocovariances. The fitted offset $2\omega^2$ is then subtracted from $m(2,\Delta)$ and the slope is re-estimated. %An alternative that estimates everything at once, a joint nonlinear fit of $H$, $\kappa$, and the offset, was developed and evaluated in simulation but because $H$ and the mean-reversion parameter

\subsection{Mean-reversion contamination of the second moment}\label{sec:method:mrc}

Volatility is not exactly a fractional Brownian motion, and volatility models automatically include mean reversion. The simplest and most tractable model that combines both features is the fractional Ornstein-Uhlenbeck (fOU) process \citep{comte1998long, cheridito2003fractional, gatheral2018volatility}. This subsection quantifies exactly how mean reversion contaminates the second-moment slope, for every $H\in(0,1)$, and provides a bias formula. \citet{garcin2022long} document that mean reversion bends the log-log plot at long lags. The large-lag autocovariance decay of order $\Delta^{2H-2}$ is classical \citep{cheridito2003fractional}, and the $\Delta^2$ order of the short-lag covariance remainder appears as an error bound in the multivariate fOU treatment of \citep{dugo2026multivariate}. The proposition below gives the law itself, with explicit constant $(1-H)\Gamma(1+2H)$ valid on all of $H\in(0,1)$. The proposition below expands the local slope of $M_2(\Delta)$ at small $\kappa\Delta$.

Let $X$ be a stationary fOU process,
\begin{align*}
dX_t = -\kappa X_tdt + \nu dB_t^H, \qquad \kappa > 0, \ H \in (0,1),
\end{align*}
with second moment $M_2(\Delta) = \mathbb{E}[(X_{t+\Delta}-X_t)^2]$ and local log-log slope
\begin{align*}
	\alpha(\Delta) = \frac{\partial \log(M_2(\Delta))}{\partial \log(\Delta)}.
\end{align*}
Without mean reversion $M_2(\Delta) = \nu^2\Delta^{2H}$ and $\alpha \equiv 2H$. The following proposition quantifies the deviation induced by $\kappa > 0$.

\begin{proposition}[Mean-reversion effect on the second moment]
\label{prop:mrc}
For every $H\in(0,1)$, as $\kappa\Delta \to 0$,
\begin{align}\label{eq:mrc}
\alpha(\Delta) = 2H - (1-H)\Gamma(1+2H)(\kappa\Delta)^{2-2H} + o\left((\kappa\Delta)^{2-2H}\right).
\end{align}
\end{proposition}
\textbf{Remark.} The proof, based on the spectral representation of the stationary fOU process and a Mellin-type evaluation of the resulting integral, is in \ref{app:mrc}.

The exponent is $2-2H$ for \emph{all} $H$. Since $2-2H > 1$ whenever $H < 1/2$, the contamination is \emph{sub-linear} in $\kappa\Delta$ for rough processes, i.e., mean reversion distorts the scaling less than in the diffusive case. And at $H=1/2$, where $2-2H = 1$ and $(1-H)\Gamma(1+2H) = 1/2$, the proposition recovers the exact Ornstein-Uhlenbeck result:
\begin{corollary}[$H=1/2$]\label{cor:ou-exact}
For $H=1/2$, $M_2(\Delta) = \tfrac{\nu^2}{\kappa}(1-e^{-\kappa\Delta})$ and 
\begin{align*}
	\alpha(\Delta) = \frac{\kappa\Delta}{e^{\kappa\Delta}-1}=1-\frac{\kappa\Delta}{2} + \frac{(\kappa\Delta)^2}{12} + O\left((\kappa\Delta)^4\right),
\end{align*}
whose leading correction $-\kappa\Delta/2$ coincides with \eqref{eq:mrc}.
\end{corollary}

Integrating \eqref{eq:mrc} gives the shape of the contaminated second moment,
\begin{align*}
	\log(M_2(\Delta) = \text{const} + 2H\log(\Delta) - \frac{\Gamma(1+2H)}{2}(\kappa\Delta)^{2-2H} + o(.),
\end{align*}
so that the ordinary least-squares slope computed on lags $\Delta = 1,\ldots, \Delta_{\max}$ satisfies
\begin{align}
	\mathbb{E}[\Hhat] \simeq H -\frac{\Gamma(1+2H)}{4}\kappa^{2-2H}\sum_{\Delta=1}^{\Delta_{\max}}w_{\Delta}\Delta^{2-2H}, \qquad w_\Delta = \frac{\log(\Delta)-\overline{\log(\Delta)}}{\sum_{\Delta'=1}^{\Delta_{\max}}\left(\log(\Delta')-\overline{\log(\Delta)}\right)^2},
\end{align}
where $\overline{\log(\Delta)}$ is the grid average and $w_\Delta$ are the least-squares weights that extract a slope on the log-log grid. Table \ref{tab:mrc-bias} reports the exact mean estimate over the empirically relevant range, computed by numerical integration of the fOU second moment. Section \ref{sec:simulation} reproduces these values in simulations. At $\Delta_{\max} = 10$ the distortion is at most $0.005$ in $H$ for $\kappa \leq 0.02$, an order of magnitude below the effects the paper reports. It grows by roughly a factor of five at $\Delta_{\max} = 40$ and by a factor of 20 at $\Delta_{\max} = 100$. On a short-lag window, mean-reversion contamination is proved negligible. 

\begin{table}[t]
\centering
\caption{Mean estimate $\E[\Hhat]$ under mean reversion, for ordinary least squares on lags
$\Delta=1,\dots,\Delta_{\max}$ computed by numerical integration of the exact fractional Ornstein-Uhlenbeck second moment.}
\label{tab:mrc-bias}
\begin{tabular}{llrrrr}
\toprule
& & \multicolumn{4}{c}{$\kappa$} \\
\cmidrule(lr){3-6}
$H$ & $\Delta_{\max}$ & $0.003$ & $0.010$ & $0.020$ & $0.035$ \\
\midrule
$0.10$ & $10$ & $0.0999$ & $0.0993$ & $0.0979$ & $0.0951$ \\
 & $20$ & $0.0997$ & $0.0982$ & $0.0951$ & $0.0893$ \\
 & $40$ & $0.0993$ & $0.0956$ & $0.0887$ & $0.0777$ \\
 & $100$ & $0.0974$ & $0.0864$ & $0.0704$ & $0.0519$ \\
\midrule
$0.15$ & $10$ & $0.1498$ & $0.1488$ & $0.1467$ & $0.1426$ \\
 & $20$ & $0.1496$ & $0.1473$ & $0.1426$ & $0.1344$ \\
 & $40$ & $0.1489$ & $0.1435$ & $0.1336$ & $0.1180$ \\
 & $100$ & $0.1461$ & $0.1304$ & $0.1077$ & $0.0810$ \\
\midrule
$0.20$ & $10$ & $0.1997$ & $0.1982$ & $0.1952$ & $0.1896$ \\
 & $20$ & $0.1993$ & $0.1960$ & $0.1896$ & $0.1786$ \\
 & $40$ & $0.1983$ & $0.1908$ & $0.1775$ & $0.1569$ \\
 & $100$ & $0.1944$ & $0.1733$ & $0.1431$ & $0.1068$ \\
\bottomrule
\end{tabular}
\end{table}

\subsection{Hurst parameter from option prices}
\label{sec:method:implied}

Under rough volatility, the at-the-money implied skew $\psi(T) = \partial_k\sigma_{\text{BS}}(k, T)\mid_{k=0}$ follows the power law
\begin{align}\label{eq:alos_fukasawa}
|\psi(T)| \sim c T^{H-1/2}, \quad T\rightarrow 0,
\end{align}
as reported by \citep{alos2007short, fukasawa2017short, bayer2016pricing}. So the slope of $\log(|\psi(T)|)$ on $\log(T)$ recovers $H-1/2$. In our analysis, the ATM skews are computed per (date, expiry) as described in Section \ref{sec:data}.\
\
Two estimates are reported. The first is the daily regression
% the statistics based on daily regressing $\log(|\psi(T)|)$ on $\log(T)$ through
\begin{align}\label{eq:reg:time}
\log(|\psi{t, T}|) = a_t + \beta_t \log(T) + u_{t,T}, %\text{ for all } t
\end{align}
estimated by date, which yields a time series $\hat{H}_t = \hat{\beta}_t + \tfrac{1}{2}$. We report the median of the time series per underlying, which is robust to dates with sparse maturity grids. The second imposes a common slope across dates while leaving the intercepts $a_t$ free,
\begin{align}\label{eq:reg:pooled}
\log(|\psi{t, T}|) = a_t + \beta \log(T) + u{t,T},
\end{align}
estimated in a single least-squares fit per underlying, and we report $\hat{H} = \hat{\beta} +\tfrac{1}{2}$. Since the intercepts absorb the level of the skew on each date, $\hat{\beta}$ is identified only by the variation of the skew across maturities within a date, in line with the power law \eqref{eq:alos_fukasawa} where skew levels never enter.\footnote{Equation \eqref{eq:reg:pooled} is a panel regression with date effects. Its least-squares slope aggregates the daily slopes $\hat{\beta}t$ from \eqref{eq:reg:time} by a weighted average of the daily slopes where each date is weighted by the dispersion of its log-maturity grid. We compute standard errors treating dates as independent units.} Alongside $\hat{H}$ we report the $R^2$ from the regression in \eqref{eq:reg:pooled}, the explained variance of $\log(|\psi{t, T}|)$ by log-maturity, per underlying asset, which also serves as an identification diagnostic. A daily time series of $H$ from \eqref{eq:reg:time} is also reported.

%% file: sections/05_simulation.tex
\section{Simulation Validation}\label{sec:simulation}
\subsection{Design}\label{sec:simulation:design}

We use a stationary fOU log-variance with Hurst parameter and mean reversion on the grid
\begin{align*}
H \in \{0.05,\, 0.10,\, 0.15,\, 0.30,\, 0.50\}, \qquad
\kappa \in \{0,\, 0.003,\, 0.010,\, 0.020,\, 0.035\},
\end{align*}
which spans half-lives from infinity to 20 days. The parameters are calibrated to SPY. The noise-adjusted dispersion of its daily log realized variance gives $\mathrm{std}(X) = 1.05$ with mean annualized volatility of about $12.8\%$. Fitting the fOU second moment to its empirical counterpart at long lags gives $\hat{\kappa}=0.009$, and its raw and corrected estimates ($0.19$ and $0.25$), which motivates a base value $H=0.2$. The benchmark cell is therefore $(H, \kappa) = (0.2, 0.010)$ yielding the grid provided. For $\kappa=0$, one obtains the classical fBm with the exact power law \eqref{eq:m2_scaling} for every lag. Any window bias that appears only for $\kappa > 0$ would be due to the effect of mean reversion. 

We build the simulation paths following three steps. Fractional Gaussian noise is simulated by circulant embedding \cite{wood1994simulation}. 
%The recursion and calibration are in Appendix \ref{app:simulation}. 
We simulate 1000 paths, and each path has ten years, $T=2520$ days of 390 one-minute returns with variance $\sigma_t^2/390$ to match the granularity of the data. Finally, Gaussian microstructure noise of standard deviation $\varpi$ is added to the log-price. Its baseline level, $\varpi=10^{-4}$. A second level, $\varpi = 5\times 10^{-4}$, appearing in the literature \citep{hansen2006realized} is also tested. Five-minute returns are subsamples of the one-minute grid.

Table \ref{tab:sim_ols} applies the regression to the true log-variance $log(\sigma_t^2)$ and, in the columns with estimation error, to the log of the five-minute realized variance computed from observed returns. The estimator and correction results of Sections \ref{sec:sim:estimators} and \ref{sec:sim:correction} use the same design as the benchmark cell. Both the primary window ($\Delta_{\max} = 10$) and the long window ($\Delta_{\max} = 40$) are reported against the truth.
%\subsection{Short lag regression}
%\label{sec:sim:lag}

\begin{table}[!htbp]
\centering
\caption{Mean estimate $\Hhat$ of the lag-restricted estimator;
the truth is the row label $H$. The columns without estimation error apply
the regression to the true log-variance ($\kappa=0$: pure fBm); the
columns with estimation error apply it to the log of the five-minute
realized variance computed from the simulated one-minute returns at the
baseline noise level. $M=1,000$ paths of ten years each per cell.}
\label{tab:sim_ols}
\small
\begin{tabular}{ccrrrr}
\toprule
& & \multicolumn{2}{c}{true log-variance} & \multicolumn{2}{c}{five-minute RV} \\
\cmidrule(lr){3-4}\cmidrule(lr){5-6}
$H$ & $\kappa$ & lag 10 & lag 40 & lag 10 & lag 40 \\
\midrule
0.05 & 0     & $0.049$ & $0.049$ & $0.047$ & $0.047$ \\
0.05 & 0.003 & $0.050$ & $0.050$ & $0.048$ & $0.048$ \\
0.05 & 0.010 & $0.049$ & $0.047$ & $0.048$ & $0.045$ \\
0.05 & 0.020 & $0.049$ & $0.043$ & $0.047$ & $0.042$ \\
0.05 & 0.035 & $0.047$ & $0.036$ & $0.045$ & $0.035$ \\
0.10 & 0     & $0.099$ & $0.100$ & $0.093$ & $0.095$ \\
0.10 & 0.003 & $0.100$ & $0.099$ & $0.094$ & $0.094$ \\
0.10 & 0.010 & $0.099$ & $0.095$ & $0.094$ & $0.091$ \\
0.10 & 0.020 & $0.098$ & $0.088$ & $0.094$ & $0.085$ \\
0.10 & 0.035 & $0.094$ & $0.076$ & $0.091$ & $0.074$ \\
0.15 & 0     & $0.150$ & $0.150$ & $0.136$ & $0.139$ \\
0.15 & 0.003 & $0.150$ & $0.148$ & $0.136$ & $0.138$ \\
0.15 & 0.010 & $0.149$ & $0.143$ & $0.139$ & $0.136$ \\
0.15 & 0.020 & $0.146$ & $0.133$ & $0.139$ & $0.127$ \\
0.15 & 0.035 & $0.142$ & $0.117$ & $0.136$ & $0.112$ \\
0.20 & 0     & $0.199$ & $0.199$ & $0.171$ & $0.178$ \\
0.20 & 0.003 & $0.199$ & $0.197$ & $0.173$ & $0.178$ \\
0.20 & 0.010 & $0.198$ & $0.190$ & $0.181$ & $0.178$ \\
0.20 & 0.020 & $0.195$ & $0.178$ & $0.182$ & $0.169$ \\
0.20 & 0.035 & $0.189$ & $0.157$ & $0.179$ & $0.151$ \\
0.30 & 0     & $0.299$ & $0.299$ & $0.209$ & $0.239$ \\
0.30 & 0.003 & $0.300$ & $0.297$ & $0.217$ & $0.242$ \\
0.30 & 0.010 & $0.297$ & $0.285$ & $0.249$ & $0.256$ \\
0.30 & 0.020 & $0.292$ & $0.267$ & $0.259$ & $0.247$ \\
0.30 & 0.035 & $0.283$ & $0.239$ & $0.259$ & $0.225$ \\
0.50 & 0     & $0.499$ & $0.498$ & $0.128$ & $0.233$ \\
0.50 & 0.003 & $0.497$ & $0.489$ & $0.144$ & $0.247$ \\
0.50 & 0.010 & $0.489$ & $0.469$ & $0.269$ & $0.342$ \\
0.50 & 0.020 & $0.480$ & $0.441$ & $0.334$ & $0.363$ \\
0.50 & 0.035 & $0.465$ & $0.401$ & $0.368$ & $0.352$ \\
\bottomrule
\end{tabular}
\end{table}

\subsection{Estimator comparison}
\label{sec:sim:estimators}
Table \ref{tab:sim_noise} tests the benchmark case ($(H, \kappa) = (0.2, 0.010)$) through the seven estimators in Section \ref{sec:method:rv}, implemented identically to the empirical pipeline, at both returns noise levels. Because the true integrated variance is known, the measurement error $\epsilon_t = \log(\hat{\sigma}_t^2) - \log(\mathrm{IV}_t)$ of each estimator is observable, and its variance $\omega^2$ is reported alongside the resulting estimate. Two results carry over to the data. First, the noise in the volatility measure attenuates the slope. TSRV, RK, and $\RV^{5m}$ cluster at $\omega^2$ of $0.026$ to $0.032$ with $\Hhat$ near $0.18$, and the pre-averaged estimators give lower estimates ($0.161$ to $0.167$) with larger error ($0.058$ to $0.071$). However, BPV and C-TRV carry the largest measured $\omega^2$ and still the highest estimates. Second, the noise level reorders the estimators. At the conventional level, the noise-robust estimators win. Modern liquid stocks with one-minute data thus benefit from noise-robust estimators. The empirical section benefits from this regime for quality stocks.

\begin{table}[!htbp]
\centering
\caption{Measurement error and uncorrected $\Hhat$ (lag 10) per realized
measure at $\varpi = 10^{-4}$ and $\varpi = 5\times10^{-4}$ price-noise levels. Benchmark cell
$(H,\kappa)=(0.20, 0.010)$, $T=2,520$ days, $M=1,000$ paths.
$\omega^2 = \Var(\log(\hat\sigma^2_t)-\log(\mathrm{IV}_t))$ is measured
against the known integrated variance.}
\label{tab:sim_noise}
\small
\begin{tabular}{lcccc}
\toprule
& \multicolumn{2}{c}{at $\varpi = 10^{-4}$} & \multicolumn{2}{c}{ at $\varpi = 5\times10^{-4}$} \\
\cmidrule(lr){2-3}\cmidrule(lr){4-5}
Estimator & $\omega^2$ & $\Hhat$ & $\omega^2$ & $\Hhat$ \\
\midrule
TSRV  & 0.026 & 0.183 & 0.126 & 0.143 \\
RK    & 0.031 & 0.180 & 0.064 & 0.170 \\
RV5m  & 0.032 & 0.181 & 0.337 & 0.145 \\
PAV   & 0.058 & 0.167 & 0.069 & 0.164 \\
C-TRV & 0.069 & 0.191 & 0.717 & 0.137 \\
PABPV & 0.071 & 0.161 & 0.081 & 0.158 \\
BPV   & 0.077 & 0.190 & 0.761 & 0.118 \\
\bottomrule
\end{tabular}
\end{table}

\subsection{Two-estimator correction}\label{sec:sim:correction}
%\label{sec:sim:mrc}

The correction of Section \ref{sec:method:correction} is validated on the same simulations where $\RV^{5m}$ is paired with each other estimator. Table \ref{tab:sim_pair} reports the results by pairing. For the noise-robust pairings, the correction increases the $H$ estimate. Raw and corrected estimates bracket the true (with RK, $0.180$ raw against $0.278$ corrected at lag 10 and $0.238$ at lag 40, around the true of $0.20$). The lag-10 refit overshoots while the lag-40 refit is tighter. The TSRV's slow scale is the offset-averaged five-minute realized variance, so its serially uncorrelated error component nearly coincides with that of $\RV^{5m}$. RK, whose construction is independent of the five-minute grid, is the pairing used in the empirical work. The same fit also corrects $\RV^{5m}$ itself, through $\omega_b^2 = \omega_a^2-\delta_m/2$. The same series, corrected through any of the four noise-robust pairings, returns the same answer: $0.213$ to $0.218$ at lag 10 and $0.199$ to $0.202$ at lag 40, at a truth of $0.20$ (not reported in the table). For the jump-robust pairings, the assumption of serially uncorrelated error fails. Their error is dominated by the volatility-dependent component documented in Table \ref{tab:sim_noise}, their measured $\delta_m$ flips sign against $2(\omega_a^2-\omega_b^2)$ (for C-TRV, $-0.346$ against $+0.074$), their corrected estimates barely move ($0.191$ to $0.198$), and the $\omega_b^2$ they imply for $\RV^{5m}$ absorbs the state-dependent gap and over-subtracts. The $\delta_m$-consistency diagnostic doubles as a misspecification check. The optimization converged on every path-pair combination. In the empirical sections, we therefore report raw and corrected estimates together and read them as brackets for $H$.

\begin{table}[!htbp]
\centering
\caption{Two-estimator correction by choice of the second estimator paired with $\RV^{5\mathrm{m}}$; benchmark cell at the baseline noise level, true $H = 0.20$. %``Expected'' is $2(\omega_a^2-\omega_b^2)$ from the measured 
noise variances.}
\label{tab:sim_pair}
\small
\label{tab:sim_pair}
\small
\begin{tabular}{lrrrrrr}
\toprule
Pairing & $\Hhat$ raw & $\Hhat$ corr (10) & $\Hhat$ corr (40) \\
\midrule
C-TRV  & 0.191 & 0.198 & 0.190 \\
BPV   & 0.190 & 0.195 & 0.188 \\
RK    & 0.180 & 0.278 & 0.238 \\
TSRV  & 0.183 & 0.291 & 0.245 \\
PAV   & 0.167 & 0.287 & 0.242 \\
PABPV & 0.161 & 0.287 & 0.242 \\
\bottomrule
\end{tabular}
\end{table}

%\begin{table}[!htbp]
%\centering
%\caption{Two-estimator correction by choice of the second estimator paired with
%$\RV^{5\mathrm{m}}$; benchmark cell at the baseline noise level, true
%$H = 0.20$. %``Expected'' is $2(\omega_a^2-\omega_b^2)$ from the measured
%noise variances.
%}
%\label{tab:sim_pair}
%\small
%\label{tab:sim_pair}
%\small
%\begin{tabular}{lrrrrrr}
%\toprule
%Pairing & $\delta_m$ & expected & $\Hhat$ raw & $\Hhat$ corr (10) & $\Hhat$ corr (40) \\
%\midrule
%C-TRV & $-0.346$ & $+0.074$ & 0.191 & 0.198 & 0.190 \\
%BPV   & $-0.359$ & $+0.090$ & 0.190 & 0.195 & 0.188 \\
%RK    & $+0.122$ & $-0.003$ & 0.180 & 0.278 & 0.238 \\
%TSRV  & $+0.143$ & $-0.013$ & 0.183 & 0.291 & 0.245 \\
%PAV   & $+0.193$ & $+0.051$ & 0.167 & 0.287 & 0.242 \\
%PABPV & $+0.218$ & $+0.077$ & 0.161 & 0.287 & 0.242 \\
%\bottomrule
%\end{tabular}
%\end{table}

\begin{figure}[t]
\centering
\includegraphics[width=0.72\textwidth]{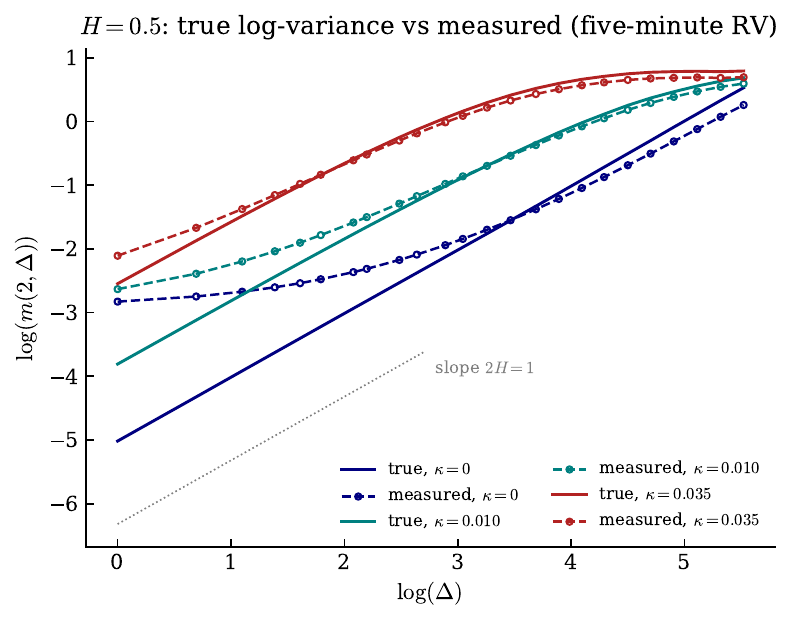}
\caption{Scaling of the second moment at $H = 0.5$ for $\kappa \in \{0, 0.010, 0.035\}$. Second moment from true log-variance, with the short-lag slope $2H = 1$ (grey dotted line) and the mean-reversion bend at long lags for $\kappa > 0$. Second moment is measured from five-minute realized variance on the simulated returns (dashed with markers). %The measurement offset pins the curves from below and flattens the short lags, most severely at $\kappa = 0$, which is what the lag-restricted regression reads as extreme roughness.
}
\label{fig:smooth_scaling}
\end{figure}

Table \ref{tab:sim_corr_grid} extends the RK-based correction across the $(H, \kappa)$ grid, from the rough regime through the diffusive benchmark $H=1/2$. The measured $\omega^2$ is flat across cells at approximately $0.031$, matching the empirical $RK$ value. The raw estimate sits below the truth and the corrected lag-10 estimate at or above it, for all $H\leq 0.3$ and every $\kappa$, with the corrected value overshooting most at the roughness cells ($0.146$ at a truth of $0.05$), so the corrected lag-10 estimate is an upper edge of the bracket, not a point estimate. The corrected lag-40 estimate is the tighter of the two refits. At the calibrated $(0.20, 0.010)$ it reads $0.236$ against a raw $0.180$ and a truth of $0.20$. The smooth rows quantify a regime the design fails at, most severely at $\kappa = 0$. There, the log-variance is a pure random walk, yet the estimation provides a rough estimate. Figure \ref{fig:smooth_scaling} shows the mechanism. The true log-variance scales with slope $2H = 1$ at short lags for every $\kappa$, while the measured second moment runs nearly flat over the first set of lags, flattest at $\kappa = 0$. 

The two regimes leave different fingerprints. Under a rough truth, raw and corrected estimates are close and bracket the truth; the two corrected refits sit close together; cleaner measurement moves the estimate towards the true value; and the estimate is stable across windows. Under smooth true $H$, the correction leaves a large unexplained gap; its two refits disagree in the opposite direction, $0.29$ at lag 10 against $0.38$ at lag 40 for $\kappa = 0$. The empirical estimates in Sections \ref{sec:results:equities} and \ref{sec:results:cross} show tight corrected brackets, agreeing refits, window stability, and saturation well below one half.

\begin{table}[!htbp]
\centering
\caption{Two-estimator correction (RK paired with
$\RV^{5\mathrm{m}}$) across the $(H,\kappa)$ grid at the baseline noise
level; mean estimates, truth in the row label, $B = 400$ paths per cell.
The $\omega^2$ column is the measured error variance of RK. Cells
with $H \le 0.20$ use the empirical rough-side bound $H \in (0, 0.49)$;
higher cells lift it to $0.99$. The optimization converged on every path.}
\label{tab:sim_corr_grid}
\small
\begin{tabular}{ccrrrr}
\toprule
$H$ & $\kappa$ & $\omega^2$ & $\Hhat$ raw (10) & $\Hhat$ corr (10) & $\Hhat$ corr (40) \\
\midrule
0.05 & 0     & $0.031$ & $0.048$ & $0.146$ & $0.110$ \\
0.05 & 0.010 & $0.031$ & $0.048$ & $0.158$ & $0.108$ \\
0.05 & 0.035 & $0.031$ & $0.045$ & $0.165$ & $0.099$ \\
0.10 & 0     & $0.032$ & $0.094$ & $0.198$ & $0.160$ \\
0.10 & 0.010 & $0.031$ & $0.093$ & $0.209$ & $0.162$ \\
0.10 & 0.035 & $0.031$ & $0.091$ & $0.203$ & $0.136$ \\
0.15 & 0     & $0.032$ & $0.136$ & $0.268$ & $0.218$ \\
0.15 & 0.010 & $0.031$ & $0.139$ & $0.236$ & $0.196$ \\
0.15 & 0.035 & $0.031$ & $0.135$ & $0.219$ & $0.160$ \\
0.20 & 0     & $0.033$ & $0.170$ & $0.305$ & $0.260$ \\
0.20 & 0.010 & $0.031$ & $0.180$ & $0.277$ & $0.236$ \\
0.20 & 0.035 & $0.031$ & $0.178$ & $0.263$ & $0.199$ \\
0.30 & 0     & $0.036$ & $0.209$ & $0.456$ & $0.372$ \\
0.30 & 0.010 & $0.031$ & $0.246$ & $0.376$ & $0.326$ \\
0.30 & 0.035 & $0.031$ & $0.257$ & $0.368$ & $0.283$ \\
0.50 & 0     & $0.075$ & $0.128$ & $0.288$ & $0.378$ \\
0.50 & 0.010 & $0.031$ & $0.261$ & $0.656$ & $0.534$ \\
0.50 & 0.035 & $0.031$ & $0.363$ & $0.677$ & $0.486$ \\
\bottomrule
\end{tabular}
\end{table}

The correction assumes the measurement error is serially uncorrelated on the log scale. Heavy price noise violates this. Its bias on realized variance is constant in level, so in logs it becomes volatility-dependent and serially correlated. This is observed at $\varpi = 5\times 10^{-4}$ in Table \ref{tab:sim_noise}. The correction therefore applies where price-level noise is small relative to signal, which the empirical noise envelope verifies for our sample. 

%% file: sections/06_results_equities.tex
\section{Results: Equities}\label{sec:results:equities}

\subsection{Cross-sectional distribution}
\label{sec:results:equities:dist}

\begin{figure}[t]
\centering
\includegraphics[width=0.85\textwidth]{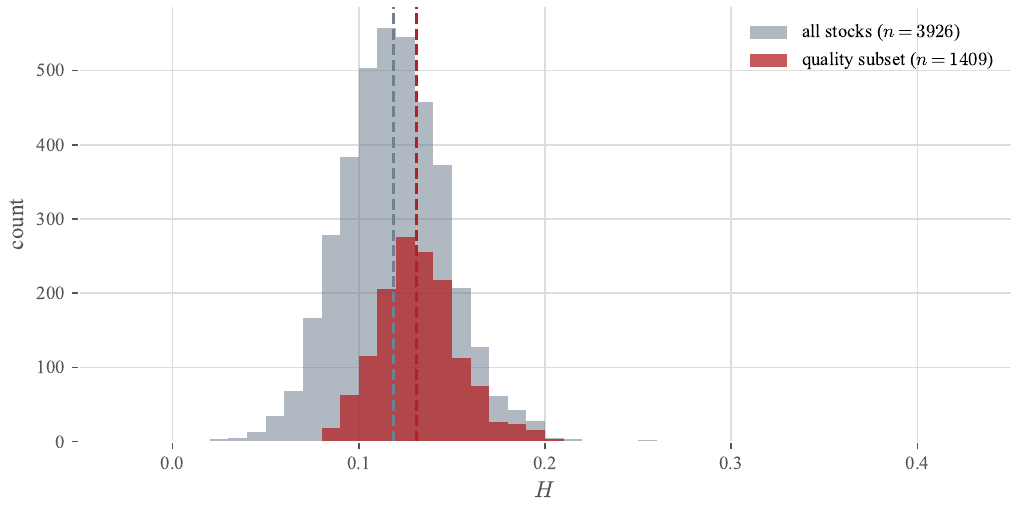}
\caption{Distribution of the realized Hurst parameter (TSRV, lag-10) across 3,926 Nasdaq stocks, all stocks, and a quality subset. Dashed vertical lines are the medians.}
\label{fig:equity_distribution}
\end{figure}

Figure \ref{fig:equity_distribution} shows the distribution of the Hurst estimate from realized volatility (TSRV) using a 10-day lag across 3,926 Nasdaq stocks. The distribution is tight and unambiguously rough. The median is $0.119$ across all assets, with an interquartile range of $[0.100, 0.137]$. When reporting a quality subset (stocks with at least 500 days and at least 80\% of traded minutes on most days), the median $H$ goes up slightly to $0.131$ with an interquartile range of $[0.117, 0.145]$ and a median R-squared of $0.988$. All values lie well below $H=0.5$.\footnote{We summarize by the medians to avoid the effect of outliers. However, the mean agrees with the median to 0.001 level for both the realized and implied Hurst estimators.}
The distribution of Figure \ref{fig:equity_distribution} shows that quality data has a higher Hurst exponent (the median is 0.012 higher). The quality subset contains 1,409 stocks with at least 500 days, at least half of which have 80\% or more of the 390 regular-session minutes traded. Measurement error biases $\hat{H}$ downward and is worst for illiquid stocks, so a small positive shift is expected. Second, the estimate is stable across all seven realized volatility estimators, spanning noise-robust, jump-robust, and the 5-min benchmark. Medians in the quality subset range from 0.114 (PAV) to 0.175 (CTRV).%\\
%\\

We also include three further checks to refine the measurement; each raises $\hat{H}$. The two-estimator correction of Section \ref{sec:method:correction} using RK and $RV^{5m}$, raises the quality subset RK median at the lag-10 window to 0.140 (from 0.125) and at the lag-40 to 0.123 (from 0.111). For the 40 most liquid stocks, we use 1-second prices and rerun the estimations. The median estimated $H$ moves to 0.186 for RK, 0.15 for TSRV, and 0.170 for PAV. Aggregating to weekly and monthly realized variance gives a median of 0.149 and 0.148. For liquid stocks, realized volatility is nearly unchanged across sampling intervals from one second to ten minutes, so price-level microstructure noise is not a constraint. We conclude that almost all estimates agree that equity volatility exhibits a Hurst parameter in roughly $[0.10, 0.17]$, far from Gaussian diffusion. % from one second to ten minutes,

\subsection{Implied versus realized Hurst}
\label{sec:results:equities:implied}

For indices, the pooled skew estimates give strong results and are concentrated $\hat{H}_{IV}=0.277$ (ES), $0.253$ (XSP), 0.243 (SPY), 0.230 (SPX), $0.225$ (YM), and $0.207$ (NQ) with $R^2_{\psi}$ between $0.21$ and $0.82$ and clustered standard errors below 0.008. These estimates are robust across estimation settings (moneyness band, strike minima, OTM-only), with a maximum spread of 0.06 per asset, and they sit close to $H\approx 0.3$ reported from option prices by \citet{livieri2018rough}. When moving the maturity window to [30-day, 1-year] instead of [14-day, 1-year], the estimates become 0.240 (ES), 0.222 (XSP), 0.214 (SPY and SPX), 0.147 (NQ), and $0.139$ (YM). The 14-day window remains our primary choice as it retains the power law. Compared with the realized estimates for the same underlyings (0.186 to 0.201 for the index futures, lag-10 TSRV), implied $H$ at the 14-day window is slightly above realized $H$ by roughly $0.02$ to $0.08$. The sign of the gap is consistent with the smoothing interpretation: option prices reflect conditional expectations of future volatility, and expectation smooths roughness.

For single stocks, the implied estimation weakens. All seven stocks have $R^2_{\psi}$ below 0.3 threshold of Figure \ref{fig:cmp_implied_realized}. Among those stocks, MSFT ($0.235$), GOOGL (0.193), and AMZN (0.151) are positive and moderate, while TSLA ($0.060$), AMD ($0.004$), AAPL (-0.016), and NVDA (-0.013) sit at or near zero with comparable $R^2_{\psi}$. Their realized $H$ estimate ranges from $0.11$ to $0.14$.

\begin{figure}[t]
\centering
\includegraphics[width=0.85\textwidth]{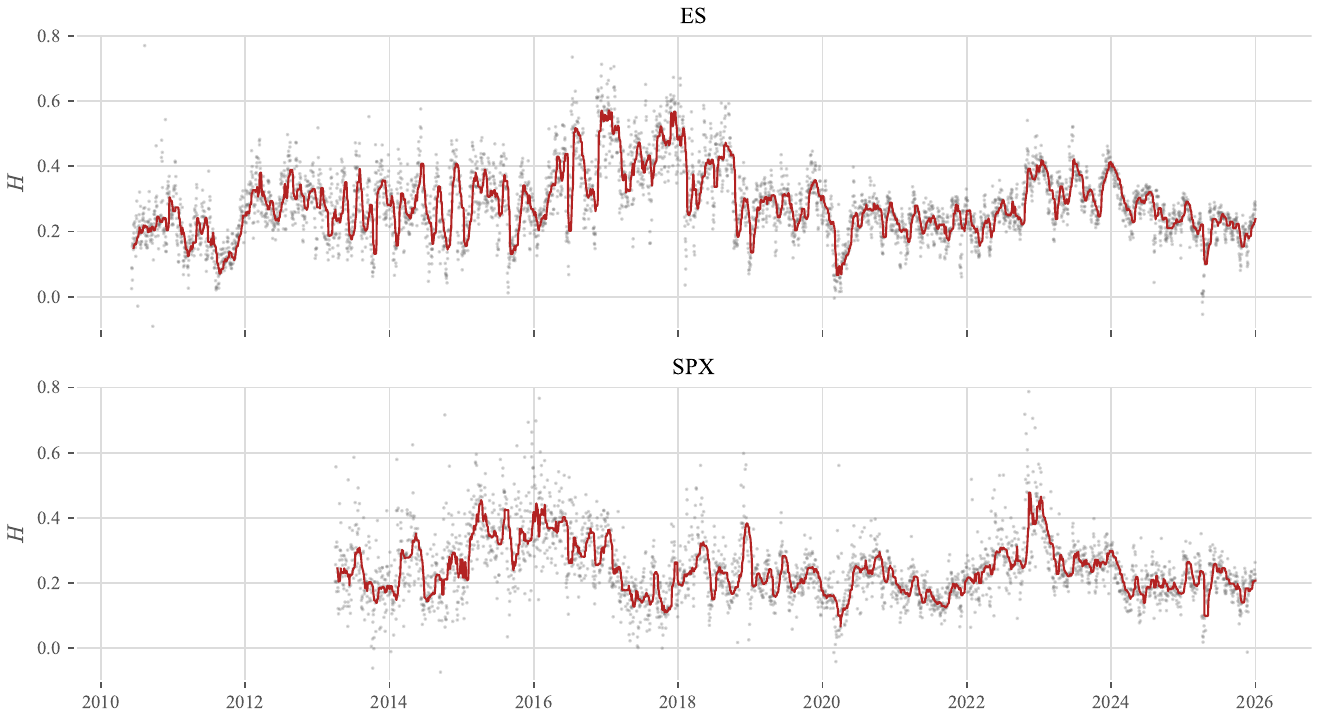}
\caption{Daily implied Hurst parameter from the Al\`{o}s/Fukasawa skew regression, ES
and SPX, unfiltered, with a 21-day rolling median.}
\label{fig:iv_timeseries}
\end{figure}

While index futures sit on the smoother side of the equity cross-section with realized $H$ around $0.2$, single stocks have a median of $0.131$. The daily implied $H$ series in Figure \ref{fig:iv_timeseries} shows a median $H$ of $0.270$ (ES) and $0.229$ (SPX), with a range between $0.07$ and $0.4$ for fifteen years.

%% file: sections/07_results_cross_asset.tex
\section{Results: Cross-Asset Patterns}\label{sec:results:cross}

\subsection{Main cross-asset finding}
\label{sec:results:cross:main}
\begin{figure}[!htbp]
\centering
\includegraphics[width=0.9\textwidth]{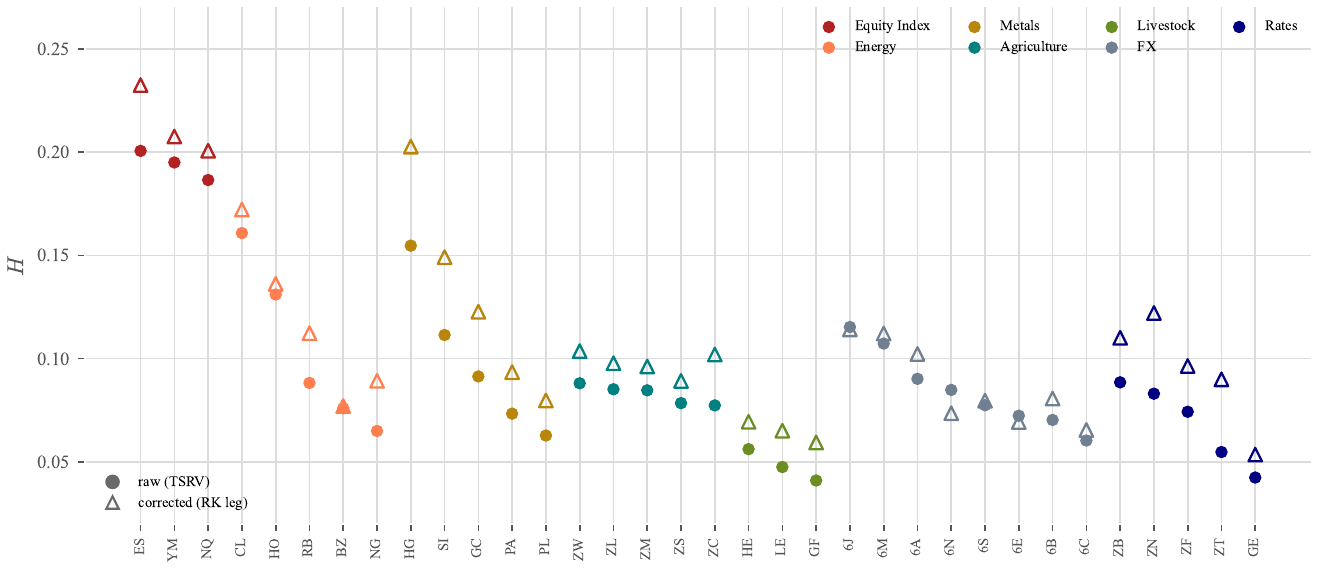}
\caption{Realized Hurst parameter across the 34 CME futures roots, colored by asset class. Filled circles: primary estimate (TSRV, lag-10). Open triangles: two-estimator noise-corrected estimate (RK leg, lag-10 refit), computed for all 34 roots; for roots below the fill screen the corrected estimate inherits the lower data quality.}
\label{fig:cross_asset}
\end{figure}

Figure \ref{fig:cross_asset} and Table \ref{tab:cross_asset} report the realized Hurst parameter (TSRV) for the 34 futures roots, grouped by asset class. On the realized side, every class has rough volatility, but in a heterogeneous way. Livestock has a median of 0.048 while equity indices have a median of 0.195, and every root is far below $H=0.5$. Rates (0.074), FX (0.081), and agriculture (0.085) have class medians that are close to each other, while values within a class spread wider (metals from 0.063 to 0.155, energy from 0.065 to 0.161). The noise-corrected $H$, shown as a triangle in Figure \ref{fig:cross_asset}, raises RK-based estimates over the 34 roots. 

\begin{table}[!htbp]
\centering
\caption{Realized Hurst parameter by asset class (TSRV, $\Delta_{\max}=10$).
Class medians with per-root estimates.}
\label{tab:cross_asset}
\footnotesize
\begin{tabular}{llc}
\toprule
Class & Roots ($\Hhat$) & Median \\
\midrule
Equity index & ES (0.201), YM (0.195), NQ (0.186) & (0.195) \\
Metals       & HG (0.155), SI (0.111), GC (0.091), PA (0.073), PL (0.063) & 0.091 \\
Energy       & CL (0.161), HO (0.131), RB (0.088), BZ (0.076), NG (0.065) & 0.088 \\
Agriculture  & ZW (0.088), ZL (0.085), ZM (0.085), ZS (0.078), ZC (0.077) & 0.085 \\
FX           & 6J (0.115), 6M (0.107), 6A (0.090), 6N (0.085),         & 0.081 \\
             & 6S (0.077), 6E (0.072), 6B (0.070), 6C (0.060)          &       \\
Rates        & ZB (0.089), ZN (0.083), ZF (0.074), ZT (0.055), GE (0.042) & 0.074 \\
Livestock    & HE (0.056), LE (0.048), GF (0.041) & 0.048 \\
\bottomrule
\end{tabular}
\end{table}

%\ref{app:rob:estimators} reports class medians for all seven estimators. 
When looking at all seven estimators, every class median stays in the rough range. BPV moves class medians by at most $0.07$ relative to TSRV, so price jumps do not drive the estimates. C-TRV is the highest estimator in every class except livestock. The cross-estimator spread ranges from $0.04$ (livestock) to $0.15$ (FX).

\subsection{Implied-realized comparison by asset class}
\label{sec:results:cross:implied}

\begin{figure}[t]
\centering
\includegraphics[width=0.7\textwidth]{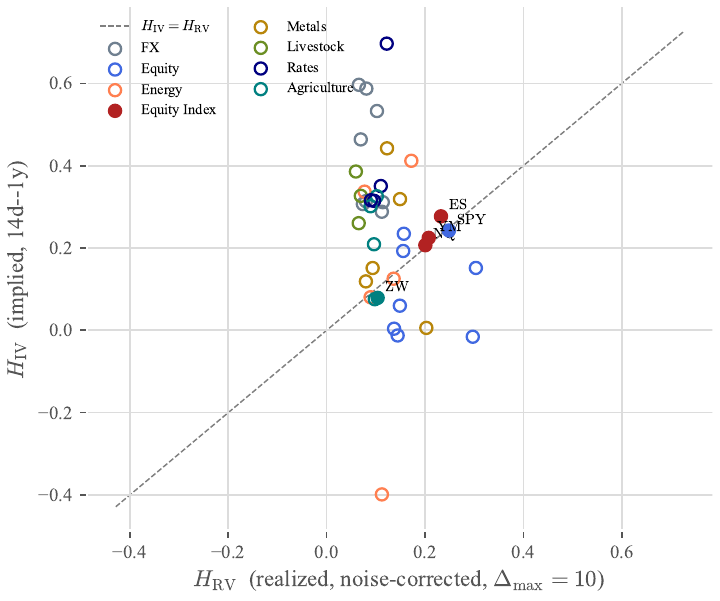}
\caption{Implied versus realized Hurst parameter (TSRV corrected lag-10), one point per asset, colored by class. Filled markers: implied $R^2_\psi \ge 0.3$ (identified). Hollow markers: weakly identified implied estimates. Dashed line: parity.}
\label{fig:cmp_implied_realized}
\end{figure}

Figure \ref{fig:cmp_implied_realized} plots implied against realized $H$ for 41 assets with both estimates, where we mark assets with $R^2_{\psi}\leq 0.3$. The picture splits the market in two. The identified set at the threshold (filled markers) is the equity index cluster, which sits near the diagonal with implied slightly above realized. Two boundary cases remain: XSP slightly below the cutoff ($R^2_{\psi} = 0.27$), and one agricultural root passes it, ZW at $0.34$, landing close to the diagonal (implied 0.080 against realized 0.088), a supporting observation from outside equities. Rates and FX fall below $0.16$, while the remaining commodities fall below $0.27$, with RB at 0.26 and $H_{IV}\approx -0.4$. Finally, ZN reports an implied $H$ of 0.69. In those markets, however, the scatter reflects the skew power law's failure to identify $H$, not necessarily volatility behavior. The cross-asset conclusions therefore rest on the realized estimates, and the implied channel is used where it is identified.

\begin{figure}[t]
\centering
\includegraphics[width=0.7\textwidth]{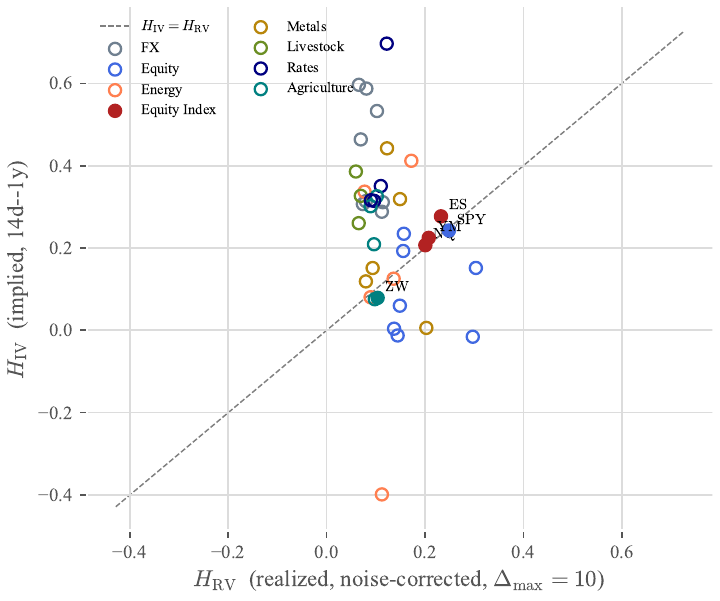}
\caption{Implied versus noise-corrected realized Hurst parameter (RK leg, lag-10 refit), one point per asset, colored by class; markers as in Figure~\ref{fig:cmp_implied_realized}.}
\label{fig:cmp_implied_realized_corr}
\end{figure}

Figure \ref{fig:cmp_implied_realized_corr} repeats the comparison with the noise-corrected realized estimate on the x-axis. For the identified assets, the correction closes most of the gaps to the diagonal. NQ and SPY move to within 0.01, the ES gap moves from 0.08 to 0.05, and ZW stays close (0.079 implied against 0.104 corrected). The correction removes a known downward bias, so the corrected panel is the more direct comparison, with option data reporting roughness similar to the realized measure.

%% file: sections/08_failure_taxonomy.tex
\section{Failure Taxonomy}\label{sec:failure}

 The realized volatility framework continues to deliver rough, well-fitted estimates throughout. The implied volatility channel, however, reports two breaks. The skew regression stops identifying $H$ where there is no leverage skew (rates and FX) and where seasonality distorts the maturity grid (natural gas). A third break concerns commodities whose interclass spread of the $H$ estimate is as wide as the spread across asset classes, with no single $H$ representing them.\\
\\
%\subsection{Rates: absence of the leverage effect}
%\label{sec:failure:rates}
\textbf{Rates.} The implied skew regression identifies $H$ from the maturity decay of the at-the-money skew, and the skew exists because volatility and returns are correlated. In Treasury futures options, that correlation is close to zero. Smiles are nearly symmetric, the sign of the measured skew flips across dates and maturities, and $\log(|\psi|)$ is dominated by noise. The pooled estimates are then meaningless: $\Hhat_{IV}=0.70$ (ZN), $0.35$ (ZB), $0.32$ (ZT), $0.31$ (ZF), all with $R^2_{\psi}$ between $0.02$ and $0.03$. FX options fail the same way: near-symmetric smiles yields $R^2_{\psi}$ from $0.004$ (6A) to $0.16$ (6M) and estimates from $0.29$ to $0.60$. Unspanned stochastic volatility in fixed income \citep{trolle2009interest} describes the decoupling behind this. The realized estimates for the same contracts are roughly $0.06$ to $0.09$ for rates and $0.06$ to $0.12$ for FX, with regression $R^2$ above 0.86. The failure is specific to the implied channel.\\
\\
\textbf{Natural gas.} Seasonality does not affect the realized measure. Deseasonalizing the level from $\log$ RV changes $\hHat$ by at most $0.002$ across NG, ZC, ZS, ZW, HO, and CL. It does affect the implied measure. The NG skew term structure mixes a seasonal maturity pattern with the power-law decay that identifies $H$, so the estimate is fragile to the composition of the option universe. With monthly options, only the pooled regression gives $\Hhat_{IV}=0.26$. Adding the weekly options triples $R^2_{\psi}$ and moves the estimate to $0.080$. A composition choice that does not affect ES moves NG by $0.18$.\\
\\
\textbf{Commodity heterogeneity.} The third failure is assuming that commodities represent a single class in terms of $H$. The evidence here is that realized measures show a large spread across commodities. The implied measure is unidentified for every commodity root but ZW, so it can neither support nor contradict the class structure. Within energy, CL ($0.16$) and HO ($0.13$) sit twice as high as NG, RB, and BZ ($0.07$ to $0.09$). Within metals, HG ($0.16$) and SI ($0.11$) are against GC ($0.09$) and the platinum group ($0.06$ to $0.07$). Livestock is below every class ($ 0.04$ to $ 0.06$). The within-class spread is as large as the between-class spread and seems to correlate loosely with storage and delivery mechanics. Storable, arbitraged contracts look more like financial assets, while contracts tied to inelastic physical supply (livestock, gas) are roughest. We do not resolve the mechanism. A single commodity $H$ is not a well-defined object, and models calibrated on CL should not be transferred to $NG$ or $LE$, for example.

%% file: sections/09_conclusion.tex
\section{Conclusion}\label{sec:conclusion}

We measure volatility across 3,926 equities, 34 futures roots, and 44 option underlyings on a single pipeline. Realized volatility is rough everywhere. Class medians run from $0.05$ (livestock) through $0.08$ to $0.10$ (rates, FX, agriculture, energy, metals) to $0.13$ (single stocks) and $0.20$ (equity indices). The implied measure identifies $H$ only for the equity index class, where estimates of $0.21$ and $0.28$ sit just above realized, in line with \citet{livieri2018rough}. Cleaner measurement, through quality conditioning, the offset correction, one-second prices, or weekly aggregation, moves the quality-subset equity median (TSRV) from $0.13$ to $0.14-0.15$, and no refinement moves it towards $0.5$.

The short lag second-moment regression at $\Delta_{\max}=10$ is a simple primary estimator, and Proposition \ref{prop:mrc} justifies it. The simulation adds a caution: the joint fit of $H$, mean reversion, and noise is not identified in realistic samples and inflates $\Hhat$ in rough regimes.

Several limitations remain. The non-equity cross-sections rest on a handful of contracts per class. The fOU lens is a model choice: $H$ is a scaling exponent of log volatility, and its interpretation as fractional-driver memory is model dependent. The implied estimates depend on option data quality and, for seasonal underlyings, on the listed maturity grid. 

Two extensions follow. On the pricing side, the commodity results motivate rough forward-variance models on the futures curve, with maturity-dependent roughness and the Samuelson effect. The data pipeline extends directly to the full curve. On the measurement side, rates volatility needs an implied $H$ measure that is not affected by the absence of leverage skew, for instance, the curvature term structure or variance-swap replication.

%% file: sections/A1_mrc_proofs.tex
\section{Proofs of the mean-reversion propositions}\label{app:mrc}

The stationary fOU process $dX_t = -\kappa X_t dt + \nu dW_t^H$ has spectral density
\begin{align}\label{eq:S-lambda}
S(\lambda) = \frac{\nu^2\Gamma(2H+1)\sin(H \pi)}{2\pi}\frac{|\lambda|^{1-2H}}{\lambda^2+\kappa^2},
\end{align}
see \citet{hult2003approximating, shi2024spectral}. The second moment of the increment follows from the spectral density
\begin{align*}
	M_2(\Delta) = \mathbb{E}[(X_{t+\Delta}-X_t)^2] = 4\int_{0}^{\infty} (1-\cos(\Delta \lambda))S(\lambda)d\lambda.
\end{align*}
Substituting $u=\Delta\lambda$ we can rewrite
\begin{align*}
	M_2(\Delta) = C(H,\nu)\Delta^{2H}I(\kappa\Delta),
\end{align*}
with $C(H,\nu) = \frac{\nu^2\Gamma(2H+1)\sin(H\pi)}{2\pi}$ and where
\begin{align}\label{eq:I-epsilon}
	I(\varepsilon) = \int_{0}^{\infty}\frac{u^{1-2H}(1-\cos(u))}{u^2+\varepsilon^2}du.
\end{align}
Therefore 
\begin{align*}%\label{eq:log-M2}
	\log(M_2(\Delta)) = \log(C(H,\nu)) + 2H\log(\Delta) + \log(I(\kappa\Delta)),
\end{align*}
and
\begin{align}\label{eq:alpha-delta}
	\alpha(\Delta) = \frac{\partial \log(M_2(\Delta))}{\partial \log(\Delta)} = 2H + \frac{\varepsilon I'(\varepsilon)}{I(\varepsilon)}, \qquad \varepsilon=\kappa\Delta.
\end{align}
We then need to expand $I$ near $0$. Deriving under the integral sign, we have
\begin{align*}
	I'(\varepsilon) = -2\varepsilon\int_{0}^{\infty}\frac{u^{1-2H}(1-\cos(u))}{(u^2+\varepsilon^2)^2}du.
\end{align*}
On the other hand
\begin{align*}
	I(\varepsilon) \to_{\varepsilon \to 0} I(0) &= \int_{0}^{\infty}u^{-1-2H}(1-\cos(u))du\\
	& = \underbrace{\left[\frac{u^{-2H}}{-2H}(1-\cos(u))\right]_{0}^{\infty}}_{=0} + \frac{1}{2H}\int_{0}^{\infty}u^{-2H}\sin(u)du\\
	& = \frac{\Gamma(1-2H)}{2H}\cos(H \pi) =-\Gamma(-2H)\cos(H \pi).
\end{align*}
The last line uses the Mellin transform of the sine, $\int_{0}^{\infty}u^{\nu-1}\sin(u)du=\Gamma(\nu)\sin(\pi\nu/2)$, and using $\sin(\tfrac{\pi}{2}(1-2H))=\cos(H \pi)$. Finally, the last equality uses $\Gamma(1-2H) = -2H\Gamma(-2H)$. Note that the integral is positive by construction and therefore $-\Gamma(-2H)\cos(H \pi)$ is positive as $-\Gamma(-2H)$ and $\cos(H \pi)$ have the same sign for $H\in (0,1)$.

Writing $1-\cos(u) = \tfrac{u^2}{2} - R(u)$ where $0 \leq R(u) \leq \min\left(\tfrac{u^4}{24}, \tfrac{u^2}{2}\right)$  for all $u\geq 0$. Moreover, since $(u^2+\varepsilon^2)^2 \geq u^4$, the remainder contributes 
\begin{align*}
	2\varepsilon\int_{0}^{\infty}\frac{u^{1-2H}R(u)}{(u^2+\varepsilon^2)^2}du & \leq 2\varepsilon\left(\frac{1}{24}\int_0^1u^{1-2H}du + \frac{1}{2}\int_1^{\infty}u^{-1-2H}du\right)\\
	& = 2\varepsilon\left(\frac{1}{24(2-2H)} + \frac{1}{4H}\right) = O(\varepsilon).
\end{align*}
The main term is evaluated by a Mellin-type identity
\begin{align*}
	\int_0^{\infty}\frac{x^{s-1}}{(x^2+\varepsilon^2)^2}dx = \frac{1}{2}\Gamma\left(\frac{s}{2}\right)\Gamma\left(2-\frac{s}{2}\right)\varepsilon^{s-4}, \qquad 0 < s < 4,
\end{align*}

with $s=4-2H$, and Euler's reflection formula $\Gamma(2-H)\Gamma(H) = \frac{(1-H)\pi}{\sin(H \pi)}$, we finally have
\begin{align}\label{eq:I-prime}
	I'(\varepsilon) = -\frac{(1-H)\pi}{2\sin(H \pi)}\varepsilon^{1-2H} + O(\varepsilon).
\end{align}
Back to the main term, we have 
\begin{align*}
	\int_{0}^{\infty} \frac{u^{s-1}}{(u^2+\varepsilon^2)^2}du &\underbrace{=}_{t=\tfrac{u}{\varepsilon}} \varepsilon^{s-4}\int_{0}^{\infty}\frac{t^{s-1}}{(t^2+1)^2}dt\\
	&= \varepsilon^{s-4}\frac{1}{2}\mathrm{B}\left(\tfrac{s}{2}, 2-\tfrac{s}{2}\right)\\
	& = \frac{1}{2}\Gamma\left(\tfrac{s}{2}\right)\Gamma\left(2-\tfrac{s}{2}\right), \qquad 0 < s < 4,
\end{align*}
with $s=4-2H$, $\Gamma(2-H)\Gamma(H) = \frac{(1-H)\pi}{\sin(H \pi)}$ we have

\begin{align*}
I'(\varepsilon) = -\frac{(1-H)\pi}{2\sin(H\pi)}\varepsilon^{1-2H} + O(\varepsilon).
\end{align*}
Since $O(\varepsilon) = o(\varepsilon^{1-2H})$ for all $H \in (0,1)$, inserting all into \eqref{eq:alpha-delta}, we have
\begin{align*}
	\alpha(\Delta) & = 2H + \frac{(1-H)\pi}{2\Gamma(2H)\sin(H\pi)\cos(H\pi)}\varepsilon^{2-2H} + o(\varepsilon^{2-2H})\\
	& = 2H + \frac{(1-H)\pi}{\Gamma(2H)\sin(2H\pi)}\varepsilon^{2-2H} + o(\varepsilon^{2-2H})\\
	& = 2H - (1-H)\Gamma(1+2H)\varepsilon^{2-2H} + o((\kappa\Delta)^{2-2H}),
\end{align*}
where the last line used the reflection identity $\Gamma(-2H)\Gamma(1+2H) = -\tfrac{\pi}{\sin(2H\pi)}$.\\

\hfill $\square$

\subsection{Proof of Corollary \ref{cor:ou-exact}}\label{app:mrc:cor}

For $H=0.5$, the stationary fOU autocovariance is $\gamma(\Delta) = \tfrac{\nu^2}{2\kappa}e^{-\kappa\Delta}$, so $M_2(\Delta) = 2(\gamma(0)-\gamma(\Delta)) = \tfrac{\nu^2}{\kappa}(1-e^{-\kappa\Delta})$ and
\begin{align*}
	\alpha(\Delta) & = \frac{\Delta M_2'(\Delta)}{M_2(\Delta)}\\
	& = \frac{\kappa\Delta e^{-\kappa\Delta}}{1-e^{-\kappa\Delta}} \\
	& = \frac{\varepsilon}{e^{\varepsilon}-1} = 1 -\tfrac{\varepsilon}{2} + \tfrac{\varepsilon^2}{12} + O(\varepsilon^4).
\end{align*}
The leading correction $-\tfrac{\varepsilon}{2}$ is the same as in the the Proposition where $(1-H)\Gamma(1+2H) = \tfrac{1}{2}\Gamma(2) = \tfrac{1}{2}$ for $H = \tfrac{1}{2}$.

\hfill $\square$

%% file: sections/A4_estimator_formulas.tex
\section{Realized volatility estimators}\label{app:realized}
We begin by specifying our volatility proxies. Because the volatility process is unobserved, we estimate the Hurst exponent from realized volatility measures computed from intraday high-frequency price data. We consider seven different measures proposed in the literature and provided in this section for reference. We discuss and compare a subset of these measures in \cite{liu2015does}. In the following, we work with 1-minute OHLCV data. The log-price $p_t=\log(\text{close}_t)$ is forward filled for missing prices, and $r_t=\log\left(\frac{p_t}{p_{t-1}}\right)$ define the log-returns.
\paragraph*{\textbf{5-min realized variance (RV5m)}.}
The first class of estimators is the standard realized variance
 , defined as the sum of squared intra-daily returns. This simple estimator is the sample analog of quadratic variation. In the absence of noise, it is the nonparametric maximum likelihood estimator and is therefore efficient; see \cite{andersen2001distribution} and \cite{barndorff2002econometric}. We rely on the 5-minute returns sampling frequency as suggested by \cite{liu2015does}. 
 \begin{align*}
	\sigma^2_\text{RV5m}= \sum_{i}(p_{is}-p_{(i-1)s})^2,
\end{align*}
where $s=5$ when considering 1-minute prices and $s=300$ on the 1-second grid.

\paragraph*{\textbf{Realized Kernel (RK)}} The second class of realized measures is the realized kernel (RK) estimator, see \cite{barndorff2008designing} and \cite{barndorff2009realized}. The estimator is noise-robust and consistent under microstructure noise. 

\begin{align*}
	\sigma^2_\text{RK} = \hat{\gamma}_0 + 2\sum_{l=1}^L\kappa\left(\frac{l}{L+1}\right)\hat{\gamma}_l,
\end{align*}
where:
\begin{itemize}
	\item $L$ is a bandwidth parameter.
	\item $\hat{\gamma}_l = \sum_{j=1}^{n-l}r_jr_{j+l}$ is the sample autocovariance at lag $l$
	\item $\kappa(x)$ is the Parzen kernel:
	\begin{align*}
		\kappa(x)=\begin{cases}
			1-6x^2+6|x|^3 \quad & \text{ if } |x|\leq 0.5\\
			2(1-|x|)^3 \quad & \text{ if } 0.5 < |x| \leq 1
		\end{cases}
	\end{align*}
\end{itemize}

\paragraph*{\textbf{Two-Scale Realized Variance (TSRV)}} The next estimator is the two-scale realized variance introduced by \cite{zhang2005tale}. This estimator computes a subsampled realized variance on one or more slower time scales (in our case RV5m)  and then combines it with the realized variance calculated on a faster time scale (all returns) to correct for microstructure noise. Under certain conditions on the market microstructure noise, these estimators are consistent at the optimal rate. The implementation of this estimator is based on the formula
\begin{align*}
	\sigma_{\text{TSRV}}^2 = \frac{1}{K}\sum_{k=0}^{K-1}{(\sigma_{\text{slow}}^{(k)})}^2-\frac{\bar{n}}{n}\cdot {\sigma_{\text{fast}}^{2}},
\end{align*} 
where:
\begin{itemize}
	\item $\sigma_{\text{fast}}^{2}=\sum_{i=1}^n r_i^2$ (all returns)
	\item $\sigma_{\text{slow}}$: realized variance using every $K$-th return starting from offset $k$
	\item $\bar{n}=\frac{n-K+1}{K}$: effective degrees of freedom ratio
	\item The term $\frac{\bar{n}}{n}$ is the finite-sample bias correction
\end{itemize}

\paragraph{\textbf{Pre-Averaged Realized Variance (PAV)}} Introduced by \cite{podolskij2009estimation}, the fourth estimator is the pre-average realized variance estimator. Pre-averaging applied a kernel-like weighting function to observed returns to construct pre-averaged returns:
\begin{align*}
	\bar{Y}_i = \frac{1}{\kappa_n}\sum_{j=1}^{\kappa_n}g\left(\frac{j}{\kappa_n}\right)(p_{i+j}-p_{i+j-1})
\end{align*}
where $g(x)=\min(x,1-x)$. The PAV estimate is then computed as
\begin{align*}
	\sigma_{\text{PAV}}^2 = \frac{1}{\psi_2\kappa_n}\sum_{i}\bar{Y}_i^2-\frac{{\psi}_1}{2\theta\psi_2}\frac{1}{n}\sum_{i=1}^n r_i^2,
\end{align*}
where $\psi_1$, ${\psi}_2$, $\theta$, and $\kappa_n$ are constants.

\paragraph{\textbf{Bipower Variation (BPV)}} The fifth estimator, see \cite{barndorff2004power}, separates the continuous part of quadratic variation from jumps by replacing squared returns with the product of adjacent absolute returns. A jump enters only one factor of each product, and its diffusive neighbor is $O\left(n^{-1/2}\right)$ so jumps vanish asymptotically. This estimator is given by
\begin{align*}
	\sigma_{\text{BPV}}^2 = \frac{\pi}{2	}\sum_{i=2}^n|r_i|\cdot|r_{i-1}|.
\end{align*}

\paragraph{{Pre-Averaged Bipower Variation (PABPV)}} A jump-robust version of the PAV, this estimator uses the pre-averaged returns defined as in PAV and is given by the formula
\begin{align*}
	\sigma_{\text{PABPV}}^2 = \frac{\pi}{2\kappa_n\psi_2}\sum_i|\bar{Y}_i|\cdot |\bar{Y}_{i+\kappa_n}|.
\end{align*}
The non-overlapping blocks (spacing $\kappa_n$) ensure that consecutive bipower products are asymptotically independent.
\paragraph{\textbf{Corrected threshold realized variance C-TRV}}\
\citet{mancini2009non} handles jumps by rejection. Returns exceeding a threshold are discarded to separate the continuous diffusion from the jump component. Since truncation is biased in finite samples by discarding the tail of the continuous component, \cite{corsi2010threshold} correct this by replacing above-threshold squared returns with their expected value under Gaussianity, with the threshold at 3 diffusive standard deviations. The scale comes from the day's bipower variation. We consider a special case of \cite{corsi2010threshold}'s family of estimator which can be given by
\begin{align*}
	\sigma_{\text{CTRV}}^2= \sum_{j=1}^n\left[r_j^2\mathds{1}_{r^2_j\leq 9\vartheta^2} + \kappa \vartheta^2 \mathds{1}_{r_j^2 > 9\vartheta^2}\right],
\end{align*}
with $\vartheta^2 = \frac{\sigma_{\text{BPV}}^2}{n}$, $\kappa=1+\frac{3\Phi(3)}{1-\Phi(3)}\approx 10.849$. In fact, $\kappa\vartheta^2 = \mathbb{E}[X^2|X^2>9\vartheta^2]$ for $X \sim N(0,\vartheta^2)$.

%% file: main.bbl
\begin{thebibliography}{36}
\expandafter\ifx\csname natexlab\endcsname\relax\def\natexlab#1{#1}\fi
\providecommand{\url}[1]{\texttt{#1}}
\providecommand{\href}[2]{#2}
\providecommand{\path}[1]{#1}
\providecommand{\DOIprefix}{doi:}
\providecommand{\ArXivprefix}{arXiv:}
\providecommand{\URLprefix}{URL: }
\providecommand{\Pubmedprefix}{pmid:}
\providecommand{\doi}[1]{\href{http://dx.doi.org/#1}{\path{#1}}}
\providecommand{\Pubmed}[1]{\href{pmid:#1}{\path{#1}}}
\providecommand{\bibinfo}[2]{#2}
\ifx\xfnm\relax \def\xfnm[#1]{\unskip,\space#1}\fi
%Type = Article
\bibitem[{Alfeus and Nikitopoulos(2022)}]{alfeus2022forecasting}
\bibinfo{author}{Alfeus, M.}, \bibinfo{author}{Nikitopoulos, C.S.},
  \bibinfo{year}{2022}.
\newblock \bibinfo{title}{Forecasting volatility in commodity markets with
  long-memory models}.
\newblock \bibinfo{journal}{Journal of Commodity Markets} \bibinfo{volume}{28},
  \bibinfo{pages}{100248}.
%Type = Article
\bibitem[{Alfeus et~al.(2024)Alfeus, Nikitopoulos and
  Overbeck}]{alfeus2024implied}
\bibinfo{author}{Alfeus, M.}, \bibinfo{author}{Nikitopoulos, C.S.},
  \bibinfo{author}{Overbeck, L.}, \bibinfo{year}{2024}.
\newblock \bibinfo{title}{Implied roughness in the term structure of oil market
  volatility}.
\newblock \bibinfo{journal}{Quantitative Finance} \bibinfo{volume}{24},
  \bibinfo{pages}{347--363}.
%Type = Article
\bibitem[{Al{\`o}s et~al.(2007)Al{\`o}s, Le{\'o}n and Vives}]{alos2007short}
\bibinfo{author}{Al{\`o}s, E.}, \bibinfo{author}{Le{\'o}n, J.A.},
  \bibinfo{author}{Vives, J.}, \bibinfo{year}{2007}.
\newblock \bibinfo{title}{On the short-time behavior of the implied volatility
  for jump-diffusion models with stochastic volatility}.
\newblock \bibinfo{journal}{Finance and Stochastics} \bibinfo{volume}{11},
  \bibinfo{pages}{571--589}.
%Type = Article
\bibitem[{Andersen et~al.(2001)Andersen, Bollerslev, Diebold and
  Ebens}]{andersen2001distribution}
\bibinfo{author}{Andersen, T.G.}, \bibinfo{author}{Bollerslev, T.},
  \bibinfo{author}{Diebold, F.X.}, \bibinfo{author}{Ebens, H.},
  \bibinfo{year}{2001}.
\newblock \bibinfo{title}{The distribution of realized stock return
  volatility}.
\newblock \bibinfo{journal}{Journal of Financial Economics}
  \bibinfo{volume}{61}, \bibinfo{pages}{43--76}.
%Type = Article
\bibitem[{Barndorff-Nielsen et~al.(2008)Barndorff-Nielsen, Hansen, Lunde and
  Shephard}]{barndorff2008designing}
\bibinfo{author}{Barndorff-Nielsen, O.E.}, \bibinfo{author}{Hansen, P.R.},
  \bibinfo{author}{Lunde, A.}, \bibinfo{author}{Shephard, N.},
  \bibinfo{year}{2008}.
\newblock \bibinfo{title}{Designing realised kernels to measure the ex-post
  variation of equity prices in the presence of noise}.
\newblock \bibinfo{journal}{Econometrica} \bibinfo{volume}{76},
  \bibinfo{pages}{1481--1536}.
%Type = Article
\bibitem[{Barndorff-Nielsen et~al.(2009)Barndorff-Nielsen, Hansen, Lunde and
  Shephard}]{barndorff2009realized}
\bibinfo{author}{Barndorff-Nielsen, O.E.}, \bibinfo{author}{Hansen, P.R.},
  \bibinfo{author}{Lunde, A.}, \bibinfo{author}{Shephard, N.},
  \bibinfo{year}{2009}.
\newblock \bibinfo{title}{Realized kernels in practice: trades and quotes}.
\newblock \bibinfo{journal}{The Econometrics Journal} \bibinfo{volume}{12},
  \bibinfo{pages}{C1--C32}.
%Type = Article
\bibitem[{Barndorff-Nielsen and Shephard(2002)}]{barndorff2002econometric}
\bibinfo{author}{Barndorff-Nielsen, O.E.}, \bibinfo{author}{Shephard, N.},
  \bibinfo{year}{2002}.
\newblock \bibinfo{title}{Econometric analysis of realised volatility and its
  use in estimating stochastic volatility models}.
\newblock \bibinfo{journal}{Journal of the Royal Statistical Society Series B}
  \bibinfo{volume}{64}, \bibinfo{pages}{253--280}.
%Type = Article
\bibitem[{Barndorff-Nielsen and Shephard(2004)}]{barndorff2004power}
\bibinfo{author}{Barndorff-Nielsen, O.E.}, \bibinfo{author}{Shephard, N.},
  \bibinfo{year}{2004}.
\newblock \bibinfo{title}{Power and bipower variation with stochastic
  volatility and jumps}.
\newblock \bibinfo{journal}{Journal of Financial Econometrics}
  \bibinfo{volume}{2}, \bibinfo{pages}{1--37}.
%Type = Article
\bibitem[{Bayer et~al.(2016)Bayer, Friz and Gatheral}]{bayer2016pricing}
\bibinfo{author}{Bayer, C.}, \bibinfo{author}{Friz, P.},
  \bibinfo{author}{Gatheral, J.}, \bibinfo{year}{2016}.
\newblock \bibinfo{title}{Pricing under rough volatility}.
\newblock \bibinfo{journal}{Quantitative Finance} \bibinfo{volume}{16},
  \bibinfo{pages}{887--904}.
%Type = Article
\bibitem[{Bennedsen et~al.(2022)Bennedsen, Lunde and
  Pakkanen}]{bennedsen2022decoupling}
\bibinfo{author}{Bennedsen, M.}, \bibinfo{author}{Lunde, A.},
  \bibinfo{author}{Pakkanen, M.S.}, \bibinfo{year}{2022}.
\newblock \bibinfo{title}{Decoupling the short- and long-term behavior of
  stochastic volatility}.
\newblock \bibinfo{journal}{Journal of Financial Econometrics}
  \bibinfo{volume}{20}, \bibinfo{pages}{961--1006}.
%Type = Article
\bibitem[{Bolko et~al.(2023)Bolko, Christensen, Pakkanen and
  Veliyev}]{bolko2023gmm}
\bibinfo{author}{Bolko, A.E.}, \bibinfo{author}{Christensen, K.},
  \bibinfo{author}{Pakkanen, M.S.}, \bibinfo{author}{Veliyev, B.},
  \bibinfo{year}{2023}.
\newblock \bibinfo{title}{A {GMM} approach to estimate the roughness of
  stochastic volatility}.
\newblock \bibinfo{journal}{Journal of Econometrics} \bibinfo{volume}{235},
  \bibinfo{pages}{745--778}.
%Type = Article
\bibitem[{Cheridito et~al.(2003)Cheridito, Kawaguchi and
  Maejima}]{cheridito2003fractional}
\bibinfo{author}{Cheridito, P.}, \bibinfo{author}{Kawaguchi, H.},
  \bibinfo{author}{Maejima, M.}, \bibinfo{year}{2003}.
\newblock \bibinfo{title}{Fractional {Ornstein-Uhlenbeck} processes}.
\newblock \bibinfo{journal}{Electronic Journal of Probability}
  \bibinfo{volume}{8}, \bibinfo{pages}{1--14}.
%Type = Article
\bibitem[{Comte and Renault(1998)}]{comte1998long}
\bibinfo{author}{Comte, F.}, \bibinfo{author}{Renault, E.},
  \bibinfo{year}{1998}.
\newblock \bibinfo{title}{Long memory in continuous-time stochastic volatility
  models}.
\newblock \bibinfo{journal}{Mathematical Finance} \bibinfo{volume}{8},
  \bibinfo{pages}{291--323}.
%Type = Article
\bibitem[{Corsi et~al.(2010)Corsi, Pirino and Ren{\`o}}]{corsi2010threshold}
\bibinfo{author}{Corsi, F.}, \bibinfo{author}{Pirino, D.},
  \bibinfo{author}{Ren{\`o}, R.}, \bibinfo{year}{2010}.
\newblock \bibinfo{title}{Threshold bipower variation and the impact of jumps
  on volatility forecasting}.
\newblock \bibinfo{journal}{Journal of Econometrics} \bibinfo{volume}{159},
  \bibinfo{pages}{276--288}.
%Type = Article
\bibitem[{Daluiso et~al.(2026)Daluiso, Folgar-Came{\'a}n, Pallavicini and
  V{\'a}zquez}]{daluiso2026rough}
\bibinfo{author}{Daluiso, R.}, \bibinfo{author}{Folgar-Came{\'a}n, H.},
  \bibinfo{author}{Pallavicini, A.}, \bibinfo{author}{V{\'a}zquez, C.},
  \bibinfo{year}{2026}.
\newblock \bibinfo{title}{Rough volatility dynamics in commodity markets}.
\newblock \bibinfo{journal}{arXiv preprint arXiv:2603.26514} .
%Type = Article
\bibitem[{Dugo et~al.(2026)Dugo, Giorgio and Pigato}]{dugo2026multivariate}
\bibinfo{author}{Dugo, R.}, \bibinfo{author}{Giorgio, G.},
  \bibinfo{author}{Pigato, P.}, \bibinfo{year}{2026}.
\newblock \bibinfo{title}{The multivariate fractional {O}rnstein--{U}hlenbeck
  process}.
\newblock \bibinfo{journal}{Stochastic Processes and their Applications}
  \bibinfo{volume}{192}.
%Type = Article
\bibitem[{El~Euch et~al.(2018)El~Euch, Fukasawa and
  Rosenbaum}]{elEuch2018microstructural}
\bibinfo{author}{El~Euch, O.}, \bibinfo{author}{Fukasawa, M.},
  \bibinfo{author}{Rosenbaum, M.}, \bibinfo{year}{2018}.
\newblock \bibinfo{title}{The microstructural foundations of leverage effect
  and rough volatility}.
\newblock \bibinfo{journal}{Finance and Stochastics} \bibinfo{volume}{22},
  \bibinfo{pages}{241--280}.
%Type = Article
\bibitem[{El~Euch and Rosenbaum(2019)}]{elEuch2019characteristic}
\bibinfo{author}{El~Euch, O.}, \bibinfo{author}{Rosenbaum, M.},
  \bibinfo{year}{2019}.
\newblock \bibinfo{title}{The characteristic function of rough {Heston}
  models}.
\newblock \bibinfo{journal}{Mathematical Finance} \bibinfo{volume}{29},
  \bibinfo{pages}{3--38}.
%Type = Article
\bibitem[{Fukasawa(2017)}]{fukasawa2017short}
\bibinfo{author}{Fukasawa, M.}, \bibinfo{year}{2017}.
\newblock \bibinfo{title}{Short-time at-the-money skew and rough fractional
  volatility}.
\newblock \bibinfo{journal}{Quantitative Finance} \bibinfo{volume}{17},
  \bibinfo{pages}{189--198}.
%Type = Article
\bibitem[{Fukasawa et~al.(2022)Fukasawa, Takabatake and
  Westphal}]{fukasawa2022consistent}
\bibinfo{author}{Fukasawa, M.}, \bibinfo{author}{Takabatake, T.},
  \bibinfo{author}{Westphal, R.}, \bibinfo{year}{2022}.
\newblock \bibinfo{title}{Consistent estimation for fractional stochastic
  volatility model under high-frequency asymptotics}.
\newblock \bibinfo{journal}{Mathematical Finance} \bibinfo{volume}{32},
  \bibinfo{pages}{1086--1132}.
%Type = Article
\bibitem[{Garcin and Grasselli(2022)}]{garcin2022long}
\bibinfo{author}{Garcin, M.}, \bibinfo{author}{Grasselli, M.},
  \bibinfo{year}{2022}.
\newblock \bibinfo{title}{Long versus short time scales: the rough dilemma and
  beyond}.
\newblock \bibinfo{journal}{Decisions in Economics and Finance}
  \bibinfo{volume}{45}, \bibinfo{pages}{257--278}.
%Type = Article
\bibitem[{Gatheral et~al.(2018)Gatheral, Jaisson and
  Rosenbaum}]{gatheral2018volatility}
\bibinfo{author}{Gatheral, J.}, \bibinfo{author}{Jaisson, T.},
  \bibinfo{author}{Rosenbaum, M.}, \bibinfo{year}{2018}.
\newblock \bibinfo{title}{Volatility is rough}.
\newblock \bibinfo{journal}{Quantitative Finance} \bibinfo{volume}{18},
  \bibinfo{pages}{933--949}.
%Type = Article
\bibitem[{Hansen and Lunde(2006)}]{hansen2006realized}
\bibinfo{author}{Hansen, P.R.}, \bibinfo{author}{Lunde, A.},
  \bibinfo{year}{2006}.
\newblock \bibinfo{title}{Realized variance and market microstructure noise}.
\newblock \bibinfo{journal}{Journal of Business \& Economic Statistics}
  \bibinfo{volume}{24}, \bibinfo{pages}{127--161}.
%Type = Article
\bibitem[{Hult(2003)}]{hult2003approximating}
\bibinfo{author}{Hult, H.}, \bibinfo{year}{2003}.
\newblock \bibinfo{title}{Approximating some {V}olterra type stochastic
  integrals with applications to parameter estimation}.
\newblock \bibinfo{journal}{Stochastic Processes and their Applications}
  \bibinfo{volume}{105}, \bibinfo{pages}{1--32}.
%Type = Article
\bibitem[{Jacod et~al.(2009)Jacod, Li, Mykland, Podolskij and
  Vetter}]{jacod2009microstructure}
\bibinfo{author}{Jacod, J.}, \bibinfo{author}{Li, Y.},
  \bibinfo{author}{Mykland, P.A.}, \bibinfo{author}{Podolskij, M.},
  \bibinfo{author}{Vetter, M.}, \bibinfo{year}{2009}.
\newblock \bibinfo{title}{Microstructure noise in the continuous case: The
  pre-averaging approach}.
\newblock \bibinfo{journal}{Stochastic Processes and their Applications}
  \bibinfo{volume}{119}, \bibinfo{pages}{2249--2276}.
%Type = Article
\bibitem[{Liu et~al.(2015)Liu, Patton and Sheppard}]{liu2015does}
\bibinfo{author}{Liu, L.Y.}, \bibinfo{author}{Patton, A.J.},
  \bibinfo{author}{Sheppard, K.}, \bibinfo{year}{2015}.
\newblock \bibinfo{title}{Does anything beat 5-minute {RV}? a comparison of
  realized measures across multiple asset classes}.
\newblock \bibinfo{journal}{Journal of Econometrics} \bibinfo{volume}{187},
  \bibinfo{pages}{293--311}.
%Type = Article
\bibitem[{Livieri et~al.(2018)Livieri, Mouti, Pallavicini and
  Rosenbaum}]{livieri2018rough}
\bibinfo{author}{Livieri, G.}, \bibinfo{author}{Mouti, S.},
  \bibinfo{author}{Pallavicini, A.}, \bibinfo{author}{Rosenbaum, M.},
  \bibinfo{year}{2018}.
\newblock \bibinfo{title}{Rough volatility: Evidence from option prices}.
\newblock \bibinfo{journal}{IISE Transactions} \bibinfo{volume}{50},
  \bibinfo{pages}{767--776}.
%Type = Article
\bibitem[{Mancini(2009)}]{mancini2009non}
\bibinfo{author}{Mancini, C.}, \bibinfo{year}{2009}.
\newblock \bibinfo{title}{Non-parametric threshold estimation for models with
  stochastic diffusion coefficient and jumps}.
\newblock \bibinfo{journal}{Scandinavian Journal of Statistics}
  \bibinfo{volume}{36}, \bibinfo{pages}{270--296}.
%Type = Article
\bibitem[{Mouti(2023)}]{mouti2023rough}
\bibinfo{author}{Mouti, S.}, \bibinfo{year}{2023}.
\newblock \bibinfo{title}{Rough volatility: evidence from range volatility
  estimators}.
\newblock \bibinfo{journal}{arXiv preprint arXiv:2312.01426} .
%Type = Article
\bibitem[{Podolskij and Vetter(2009)}]{podolskij2009estimation}
\bibinfo{author}{Podolskij, M.}, \bibinfo{author}{Vetter, M.},
  \bibinfo{year}{2009}.
\newblock \bibinfo{title}{Estimation of volatility functionals in the
  simultaneous presence of microstructure noise and jumps}.
\newblock \bibinfo{journal}{Bernoulli} \bibinfo{volume}{15},
  \bibinfo{pages}{634--658}.
\newblock \DOIprefix\doi{10.3150/08-BEJ167}.
%Type = Article
\bibitem[{Shi et~al.(2024)Shi, Yu and Zhang}]{shi2024spectral}
\bibinfo{author}{Shi, S.}, \bibinfo{author}{Yu, J.}, \bibinfo{author}{Zhang,
  C.}, \bibinfo{year}{2024}.
\newblock \bibinfo{title}{On the spectral density of fractional
  {O}rnstein--{U}hlenbeck processes}.
\newblock \bibinfo{journal}{Journal of Econometrics} .
%Type = Article
\bibitem[{Trolle and Schwartz(2009)}]{trolle2009interest}
\bibinfo{author}{Trolle, A.B.}, \bibinfo{author}{Schwartz, E.S.},
  \bibinfo{year}{2009}.
\newblock \bibinfo{title}{A general stochastic volatility model for the pricing
  of interest rate derivatives}.
\newblock \bibinfo{journal}{Review of Financial Studies} \bibinfo{volume}{22},
  \bibinfo{pages}{2007--2057}.
%Type = Article
\bibitem[{Wang et~al.(2023)Wang, Xiao and Yu}]{wang2023modeling}
\bibinfo{author}{Wang, X.}, \bibinfo{author}{Xiao, W.}, \bibinfo{author}{Yu,
  J.}, \bibinfo{year}{2023}.
\newblock \bibinfo{title}{Modeling and forecasting realized volatility with the
  fractional {Ornstein--Uhlenbeck} process}.
\newblock \bibinfo{journal}{Journal of Econometrics} \bibinfo{volume}{232},
  \bibinfo{pages}{389--415}.
%Type = Article
\bibitem[{Wood and Chan(1994)}]{wood1994simulation}
\bibinfo{author}{Wood, A.T.A.}, \bibinfo{author}{Chan, G.},
  \bibinfo{year}{1994}.
\newblock \bibinfo{title}{Simulation of stationary {Gaussian} processes in
  $[0,1]^d$}.
\newblock \bibinfo{journal}{Journal of Computational and Graphical Statistics}
  \bibinfo{volume}{3}, \bibinfo{pages}{409--432}.
%Type = Article
\bibitem[{Zarhali et~al.(2025)Zarhali, Aubrun, Bacry, Bouchaud and
  Muzy}]{zarhali2025volatility}
\bibinfo{author}{Zarhali, O.}, \bibinfo{author}{Aubrun, C.},
  \bibinfo{author}{Bacry, E.}, \bibinfo{author}{Bouchaud, J.P.},
  \bibinfo{author}{Muzy, J.F.}, \bibinfo{year}{2025}.
\newblock \bibinfo{title}{Why is the volatility of single stocks so much
  rougher than that of the s\&p500?}
\newblock \bibinfo{journal}{arXiv preprint arXiv:2505.02678} .
%Type = Article
\bibitem[{Zhang et~al.(2005)Zhang, Mykland and
  A{\"\i}t-Sahalia}]{zhang2005tale}
\bibinfo{author}{Zhang, L.}, \bibinfo{author}{Mykland, P.A.},
  \bibinfo{author}{A{\"\i}t-Sahalia, Y.}, \bibinfo{year}{2005}.
\newblock \bibinfo{title}{A tale of two time scales: Determining integrated
  volatility with noisy high-frequency data}.
\newblock \bibinfo{journal}{Journal of the American Statistical Association}
  \bibinfo{volume}{100}, \bibinfo{pages}{1394--1411}.

\end{thebibliography}
